\documentclass[aps,prd,showpacs,notitlepage,nofootinbib,superscriptaddress,floatfix,showkeys,twocolumn]{revtex4-1}
\usepackage{float}
\usepackage{multirow}
\usepackage{siunitx}
\usepackage{verbatim}
\usepackage{graphicx}
\usepackage{amsmath}
\usepackage[usenames]{color}
\usepackage{amssymb}
\usepackage{subcaption}
\usepackage{subcaption}
\usepackage{placeins}
\usepackage{hyperref}
\newcommand{\be}{\begin{equation}}
\newcommand{\ee}{\end{equation}}
\newcommand{\bea}{\begin{aligned}}
\newcommand{\eea}{\end{aligned}}
\newcommand{\pr}{\partial}

\newcommand{\bse}{\begin{subequations}}
\newcommand{\ese}{\end{subequations}}

\newcommand{\bmm}{\begin{multline}}
\newcommand{\emm}{\end{multline}}

\begin{document}
%\title{Interplay of black hole hair and environment through Lyapunov exponents}
%\title{Stable-unstable trajectories and Lyapunov exponents due to spin-hair interaction in rotating spacetime}
\title{Spin--Hair Induced Chaos of Spinning Test Particles in Rotating Hairy Black Holes}
\author{Surojit Dalui}
\email{surojitdalui@shu.edu.cn}
%\author{Rajesh Karmakar}
%\email{rajesh@shu.edu.cn}
\author{Xian-Hui Ge}
\email{gexh@shu.edu.cn}
\affiliation{Department of Physics, Shanghai University, 99 Shangda Road, Shanghai, 200444, China}
\begin{abstract}
We investigate the finite-time instability of massive spinning test
particles around a rotating hairy black hole generated through
gravitational decoupling.  The particle motion is described by the full
Mathisson--Papapetrou--Dixon equations with the Tulczyjew spin
supplementary condition, and the sensitivity to initial conditions is
measured using a ZAMO-projected finite-time Lyapunov analysis.  The
hairy deformation is controlled by two parameters: $\alpha$, which
sets the deviation from Kerr, and $\beta$, which changes the radial
localization of the deformation.  We show that spin--curvature coupling
and the hairy geometry can shift the evolved orbit away from the
requested seed parameters, making the empirical orbital map essential
for interpreting the dynamics.  Small-spin and geodesic trajectories
remain close to regular behavior, whereas large-spin trajectories show
stronger finite-time growth.  A scan of the $(S,\beta)$ plane shows
that the instability does not grow monotonically, but appears in
localized regions where the particle spin and the radial profile of the
hair act cooperatively.  Thus, the hairy background does not simply
rescale the Kerr result; it reorganizes the strong-field phase-space
region sampled by spinning particles.
\end{abstract}

\maketitle

\newpage

\section{Introduction}\label{intro}
%No hair theorem suggests that the stationary black holes (BHs) are characterized by three parameters only, mass, electromagnetic charge and angular momentum. However, relaxation of some of the assumptions of the no-hair theorem, such as considering BHs to be embedded in a matter source, may generate hair. Given the possibility of the existence of such hair, it is imperative to study its consequences in the astrophysical processes.
Black holes (BHs) have long played a central role in gravitational physics, 
not merely as exotic mathematical solutions of Einstein’s equations but as 
astrophysical objects whose presence has now been confirmed through direct 
observations. The landmark detections of gravitational waves by the 
LIGO--Virgo--KAGRA Collaboration \cite{LIGOScientific:2016emj,TheLIGOScientific:2017qsa,LIGOScientific:2018mvr,LIGOScientific:2020ibl,LIGOScientific:2021usb,LIGOScientific:2025rid,KAGRA:2021vkt} and the horizon-scale images obtained by the 
Event Horizon Telescope \cite{EventHorizonTelescope:2019dse,EventHorizonTelescope:2019ggy,EventHorizonTelescope:2019ths,EventHorizonTelescope:2022wkp} have established a new era in which the near-horizon 
region of compact objects can be tested with increasing precision. At the same 
time, it is well known that various ultracompact configurations including 
boson stars \cite{Liebling:2012fv}, gravastars \cite{Mazur:2001fv}, and other BH mimickers \cite{Cardoso:2019rvt} can reproduce many 
observational signatures associated with classical BHs. This motivates the 
systematic study of spacetimes that deviate from the ideal Static Spherically Symmetric (SSS) solutions and stationary rotating solutions while remaining theoretically consistent and phenomenologically viable.

Speaking of deviations from the ideal solutions, the classical no-hair theorem states that, in general relativity, stationary
black holes are uniquely determined by their mass, angular momentum, and
charge. Nevertheless, over the past several decades it has become increasingly
clear that this uniqueness is not universal. Once additional fields or modified
gravitational sectors are introduced, black holes can support a wide variety of
non-trivial hairs. Examples include scalarised black holes in scalar–tensor and
Einstein–Gauss–Bonnet theories \cite{Silva:2017uqg, Doneva:2017bvd}, solutions with scalar \cite{Herdeiro:2014goa}, vector \cite{Fan:2016jnz} or Proca \cite{Herdeiro:2016tmi} hair,
Horndeski-inspired configurations \cite{Babichev:2017guv}, non-linear electrodynamics black holes \cite{Ayon-Beato:1998hmi}, and
spacetimes sourced by effective anisotropic matter sectors \cite{Cho:2017nhx,Kim:2019hfp}. Even within general
relativity, regular black holes, dark-matter–induced metrics, and fluid-supported
compact objects can mimic or deform the standard SSS and Kerr geometry. These many
examples demonstrate that non-trivial structure around black holes originating
from additional fields, matter distributions, or effective energy–momentum
sources can consistently alter the near-horizon geometry. 

In this broader context, the minimal geometric deformation (MGD) \cite{Ovalle:2016pwp} and
gravitational decoupling (GD) \cite{Ovalle:2017fgl} approaches provide a systematic and model-independent
framework to incorporate such additional sources into a known seed metric.  The essential idea is to begin with a standard black hole solution and
deform its mass function to account for an additional conserved source
sector, while preserving the conservation of the total energy--momentum
tensor.  
When applied to SSS \cite{Ovalle:2020kpd} or rotating systems, as shown in the gravitational-decoupling
extension to axially symmetric spacetimes \cite{Contreras:2021yxe}, this method yields a
simple but powerful way to generate a SSS or a  rotating hairy black hole without reliance
on the Newman–Janis algorithm.

%It is well known that most astrophysical compact objects possess nonzero spin \cite{Reynolds:2020jwt,Miller:2014aaa}. In modeling binary systems (e.g., BH–BH, BH–NS, or NS-NS), one often adopts a leading-order approximation in which one component is treated as a structureless point particle, and internal structure effects such as spin and tidal interactions are neglected \cite{Blanchet:2002av,Porto:2016pyg}.  
Most astrophysical compact objects possess nonzero spin
\cite{Reynolds:2020jwt,Miller:2014aaa}. In many leading-order models of
binary dynamics, one component is approximated as a structureless point
particle and internal-structure effects such as spin and tidal
interactions are neglected \cite{Blanchet:2002av,Porto:2016pyg}. With
the increasing precision of gravitational-wave and electromagnetic
observations, such corrections are becoming increasingly relevant,
especially for strong-field motion near compact objects.
%Given the remarkable success in the last decade on the precision gravitational physics observation, next order corrections are now relevant. The relevance is more in probing the strong field regime of gravity such as, while analyzing orbital motion of compact objects near the BH horizon. These particles are treated as extended bodies with multipole moments.

%A quantity that responds sharply to such modifications is the Lyapunov exponent associated with unstable circular orbits \cite{Cardoso:2008bp}. Physically, it measures the exponential rate at which two nearby trajectories diverge under a small perturbation and therefore provides a direct characterization of the local instability timescale. In black hole spacetimes, this quantity is especially useful because it is governed by the curvature of the effective radial potential at the unstable orbit. Hence, even relatively small deviations of the background geometry from the Kerr case may leave a clear imprint on the corresponding value of the Lyapunov exponent.

%The Lyapunov exponent is also important from a broader perspective. In the eikonal regime of black hole perturbation theory, the instability of circular orbits is closely related to the damping behavior of quasinormal modes \cite{Cardoso:2008bp}. Therefore, studying the Lyapunov exponent provides a useful bridge between test particle dynamics and strong-gravity observables. In this sense, it serves as a sensitive probe of how geometric deformations of the black hole spacetime may affect the physical properties of the near-horizon region.

Realistic orbiting bodies, however, are not always structureless. Compact objects generally possess intrinsic spin, and once the spin degree of freedom is taken into account, the motion deviates from geodesic behavior because of spin--curvature coupling. The appropriate framework for describing such motion is provided by the Mathisson--Papapetrou--Dixon (MPD) equations \cite{Mathisson:1937zz, Mathisson:2010opl,papapetrou1951spinning, dixon1970dynamics}. The spin of the particle changes the phase-space evolution through
spin--curvature coupling, so that nearby trajectories can separate in
a way that is not captured by geodesic motion alone. Previous studies of spinning particle motion in Schwarzschild and Kerr
spacetimes have shown that the MPD system can display complicated
phase-space behavior, including chaotic dynamics for suitable choices
of spin and orbital parameters
\cite{Suzuki:1997chaos,Hartl:2002ig,Hartl:2003survey,Han:2010tp,LukesGerakopoulos:2016hyy,Zelenka:2019nyp,Yuan:2026realisticspin}.
Related analyses of spinning particle motion, conserved quantities, and
orbital structure in black hole backgrounds provide further context for
the present work \cite{Tod:1976,Semerak:1999,Tanaka:1996ht,saijo1998gravitational}.
%These results motivate extending trajectory level MPD chaos studies beyond the vacuum Kerr geometry to rotating hairy backgrounds.

%Previous studies have shown that spinning particle motion in Schwarzschild and Kerr backgrounds can exhibit complicated phase-spacestructure, including chaotic behavior diagnosed through Poincaré sections, Lyapunov exponents, and fast Lyapunov indicators \cite{Tod:1976,Semerak:1999,Suzuki:1997by,Hartl:2002ig,Hartl:2003survey,Han:2010tp}. These results motivate extending the trajectory-level MPD analysis beyond the vacuum Kerr geometry to rotating hairy backgrounds. Consequently, the combined effect of black hole rotation, geometric hair, and particle spin may lead to nontrivial changes in strong field dynamics.

%Previous studies of spinning particle motion in Kerr and Schwarzschildspacetimes have shown that the MPD system can display complicated, and in some regions chaotic, behavior when the particle spin and the orbital parameters are sufficiently large~\cite{Tanaka:1996ht, Suzuki:1997by,saijo1998gravitational,Hartl:2003survey}. These worksprovide important motivation for extending the trajectory level MPDchaos analysis beyond the vacuum Kerr geometry. In the present work weask how this picture is modified when the central rotating black hole is dressed by gravitational hair.

Spinning particle dynamics has also been explored in several non-Kerr
and modified black hole backgrounds, including higher-curvature,
regular, quantum corrected, dark matter dressed, and other effective
geometries~\cite{Zhang:2020qew,Ladino:2023pje,Tan:2024dmhalo,
Mannobova:2025uqf,Chen:2025baz,Umarov:2025lgt,Jumaniyozov:2025irx,Jumaniyozov:2025nnj}.
These studies show that spin--curvature coupling can be sensitive to
the detailed structure of the background geometry. This provides further
motivation for examining the full MPD phase-space dynamics in the
rotating hairy geometry considered here.

A natural diagnostic of sensitivity to initial conditions is the
Lyapunov exponent \cite{Hashimoto:2016dfz,Dalui:2018qqv,Lei:2020clg}. In this work, we use a finite-time Lyapunov analysis
based on the separation of two nearby, constraint-preserving solutions
of the full MPD system. This is different from the local instability
analysis of an unstable circular orbit based on an effective potential.
Here we evolve the full phase-space variables
\begin{equation}
	y=(x^\mu,p_\mu,S_\mu),
\end{equation}
and measure the growth of a projected deviation vector between nearby
spinning particle trajectories. This procedure allows us to probe the
global trajectory-level instability of the system rather than only the
local radial instability of a circular orbit.

The aim of the present work is to determine how the combined effect of
black-hole hair and spin--curvature coupling modifies the finite-time
instability of spinning test-particle trajectories.  The
rotating hairy metric changes the strong-field curvature through the
deformed mass function
\[
\tilde m(r)=M-\alpha\frac{r}{2}
\exp\left[-\frac{r}{M-\beta/2}\right].
\]
In this form, the parameter \(\alpha\) fixes the amplitude of the
deviation from Kerr, while \(\beta\) enters through the radial scale
\(M-\beta/2\).  Thus, changing \(\beta\) does not simply increase the
hair strength monotonically; it changes the radial localization of the
deformation and therefore the region of the strong-field geometry
sampled by the orbit.  The central question is then how this localized
geometric deformation combines with the spin--curvature force in the
full MPD phase space.

The main result of our analysis is that the onset and strength of the
finite-time Lyapunov response are not controlled monotonically by either
the particle spin or the hair parameter alone.  Instead, the largest
finite-time Lyapunov exponents appear in localized regions of the
\((S,\beta)\) plane, where \(S\) is the dimensionless spin magnitude and
\(\beta\) controls the radial localization of the hairy deformation.
For the rotating hairy metric and for the parameter range considered here,
the strongest regions occur around
\[
S\simeq0.75-0.85,\qquad \beta\simeq0.2-0.7,
\]
and around
\[
S\simeq0.65-0.75,\qquad \beta\simeq1.2-1.5 .
\]
This indicates that the strongest finite-time instability arises when
the spin--curvature coupling and the radially localized hairy
deformation act cooperatively.  In this sense, the hairy background
does not merely rescale the Kerr result; it reorganizes the region of
phase space in which spinning-particle trajectories become most
sensitive to initial conditions.

To organize the numerical survey, we label the initial data by requested
orbital parameters \((r_p,e,\iota)\), together with the dimensionless
spin magnitude \(S\) and the initial spin orientation.  Since
spin--curvature coupling can shift the actual trajectory away from the
requested seed orbit, we also introduce empirical orbital parameters
extracted directly from the evolved motion.  This distinction between
requested and empirical parameters is essential for interpreting the
numerical maps: the requested parameters describe how the initial data
are generated, whereas the empirical parameters describe the actual
region of phase space sampled by the spinning trajectory.

This requested--empirical distinction also clarifies the main novelty
with respect to previous Kerr MPD studies.  In the vacuum Kerr case,
geodesic-like orbital labels provide a natural way to organize
spinning-particle initial data.  In the rotating hairy background,
however, the deformation changes the strong-field geometry in which the
spin--curvature force acts.  As a result, the evolved trajectory can
sample an empirical region of phase space that differs substantially
from the requested seed labels.  Thus, the effect of the hair is not
only to change the numerical value of \(\lambda_{\max}\), but also to
reorganize the phase-space map connecting initial-data labels to the
actual orbital geometry sampled by the spinning particle.

The paper is organized as follows. In Sec.~\ref{sec:geometry}, we
introduce the rotating hairy black hole spacetime generated through
gravitational decoupling and define the parameters used in the analysis.
In Sec.~\ref{sec:mpd}, we review the MPD equations in the spin-one-form
formulation and give the conserved quantities associated with
stationarity and axisymmetry. In Sec.~\ref{sec:initial_conditions}, we
describe the construction of constraint-preserving initial data in terms
of requested orbital parameters and define the corresponding empirical
quantities. In Sec.~\ref{sec:lyapunov_method}, we explain the
deviation-vector method used to compute finite-time Lyapunov exponents,
including the projected phase-space norm and saturation criterion. In
Sec.~\ref{sec:results}, we present the numerical results, including
requested and empirical $(r_p,\iota)$ maps, running Lyapunov exponents,
logarithmic separation plots, and the combined dependence on the
particle spin and hair parameter. Finally, Sec.~\ref{sec:conclusion}
summarizes the main conclusions and discusses possible extensions.

\section{Axis symmetric hairy black hole}\label{sec:geometry}
In the gravitational decoupling (GD) approach, one starts from a known seed solution of Einstein's equations and introduces an additional conserved source that deforms the geometry in a controlled manner. This framework was first developed in the context of spherically symmetric systems and later extended to axially symmetric spacetimes, making it possible to construct rotating hairy black holes directly, without relying on the Newman--Janis algorithm \cite{Ovalle:2017fgl,Ovalle:2020kpd,Contreras:2021yxe}. In this sense, the resulting geometry can be interpreted as a Kerr black hole dressed by an additional gravitational sector, encoded effectively in a deformed mass function.

In the present work, we consider the rotating hairy black hole spacetime in Boyer--Lindquist coordinates \cite{Contreras:2021yxe},
\begin{eqnarray}
	ds^2=&&-\left(1-\frac{2r\tilde m(r)}{\Sigma}\right)dt^2
	+\frac{\Sigma}{\Delta}\,dr^2
	+\Sigma\,d\theta^2\nonumber\\
	&&-\frac{4ar\tilde m(r)}{\Sigma}\sin^2\theta\,dt\,d\phi
	+\frac{A(r,\theta)}{\Sigma}\sin^2\theta\,d\phi^2 ,
	\label{metric_rot_hairy}
\end{eqnarray}
where
\begin{eqnarray}
	\Sigma&=&r^2+a^2\cos^2\theta,\\
	\Delta&=&r^2+a^2-2r\tilde m(r),\\
	A(r,\theta)&=&(r^2+a^2)^2-a^2\Delta\sin^2\theta .
	\label{metric_functions_rot_hairy}
\end{eqnarray}
Here the deformed mass fuction is 
\begin{equation}
	\tilde m(r)
	=
	M-\alpha\frac{r}{2}\exp\!\left[-\frac{r}{\,M-\beta/2\,}\right].
	\label{mass_function_beta}
\end{equation}
With this choice, the metric function $\Delta(r)$ takes the explicit form
\begin{equation}
    \Delta(r)
    =
    r^2+a^2-2Mr
    +\alpha r^2
    \exp\left[
        -\frac{r}{M-\beta/2}
    \right].
\label{eq:explicit_delta}
\end{equation}
Here \(M\) is the ADM mass and \(a\equiv J/M\) is the Kerr rotation
parameter.  The parameter \(\alpha\) controls the amplitude of the
deformation away from the vacuum Kerr geometry, while \(\ell\) is an
auxiliary scale entering the gravitational-decoupling construction.
Their combination
\[
\beta=\alpha\ell
\]
acts as the effective hair parameter.  In the present parametrization,
\(\beta\) appears in the exponential scale
\[
L_\beta=M-\frac{\beta}{2}.
\]
Thus, changing \(\beta\) changes the radial localization of the hairy
deformation.  The condition \(\beta<2M\) ensures \(L_\beta>0\), so that
the exponential correction is well behaved in the exterior region.

In the underlying GD construction, \(\beta\) characterizes the scale
associated with the extra gravitational sector and is directly related
to the entropy increase produced by the hair relative to the vacuum
black-hole configuration \cite{Contreras:2021yxe}.  For physically
acceptable hairy configurations satisfying the energy-condition
constraints outside the horizon, one requires
\begin{equation}
	\beta<2M .
	\label{beta_bound}
\end{equation}
This upper bound ensures that the hairy deformation remains compatible
with the existence of a regular horizon structure in the parameter range
relevant to our analysis \cite{Contreras:2021yxe}.

The role of each parameter is therefore clear.  The mass \(M\) fixes
the overall gravitational scale of the geometry.  The rotation parameter
\(a\) governs frame dragging and distinguishes prograde and retrograde
orbital behavior, just as in the Kerr spacetime.  The parameter
\(\alpha\) fixes the amplitude of the deviation away from Kerr, while
\(\beta\) controls the radial scale over which the exponential
correction is dynamically important.  This radial localization modifies
the strong-field curvature region sampled by spinning-particle
trajectories and is therefore probed below through the finite-time
Lyapunov analysis.  A similar use of \(\beta\) as the natural
deformation parameter also appears in the static regular-hairy GD
construction discussed in \cite{Contreras:2021yxe}, where it controls
the departure from Schwarzschild behavior and the horizon structure in
the spherically symmetric limit.

The horizon structure of the rotating spacetime is determined by the roots of
\begin{equation}
	\Delta(r)=0,
	\label{delta_horizon}
\end{equation}
whereas the stationary limit surface is obtained from
\begin{equation}
	g_{tt}=0
	\quad \Longrightarrow \quad
	\Sigma-2r\tilde m(r)=0.
	\label{ergosurface_eq}
\end{equation}
Using Eq.~\eqref{mass_function_beta}, this condition becomes
\begin{equation}
    \Sigma-2Mr
    +\alpha r^2
    \exp\left[
        -\frac{r}{M-\beta/2}
    \right]
    =0 .
\label{eq:explicit_stationary_limit}
\end{equation}
Compared with the vacuum Kerr case, the presence of the deformed mass function \(\tilde m(r)\) shifts both the horizon and the ergoregion. Since the MPD force depends explicitly on the background curvature,
changes in the horizon and ergoregion structure can directly affect the spin--curvature coupling experienced by the particle. In particular, the deformation is exponentially suppressed at large radial distance, so the asymptotic region remains close to Kerr, while the near-horizon region can be substantially altered. This makes the geometry especially suitable for studying how black hole hair modifies the strong field curvature sampled by spinning particle trajectories and how this modification affects the finite-time separation of nearby MPD solutions.

It is also worth emphasizing the physical interpretation of the above solution. In the language of gravitational decoupling, the solution may be interpreted as a tensor-vacuum configuration, namely an effective black hole geometry sustained by an additional decoupled gravitational sector rather than by ordinary vacuum alone. In this sense, the hairy corrections encoded in \(\tilde m(r)\) provide a model-independent parametrization of extra matter-energy content or an additional gravitational interaction surrounding the black hole. This interpretation is useful for our purposes, because it allows one to study the imprint of hair on particle dynamics without committing to a unique microscopic origin of the extra source.

We emphasize that the rotating hairy solution used here is treated as an
effective GD background rather than as a unique
microscopic model of black hole hair. The parameter $\alpha$ controls the
amplitude of the deviation away from Kerr, while the combination
$\beta=\alpha \ell$ sets the characteristic scale associated with the
additional decoupled sector. The bound $\beta<2M$ is imposed in order to
remain within the physically acceptable parameter range of the GD
construction, where the horizon structure is well defined and the
energy condition requirements outside the horizon are satisfied. Thus,
the role of the present background is to provide a controlled deformation
of the strong-field Kerr geometry. Our aim is not to identify the
microscopic origin of the hair, but to determine how such an effective
deformation modifies the spin--curvature coupling and the finite-time
separation of nearby MPD trajectories.

%In the GD language, the metric can be viewed as a ``tensor-vacuum'' configuration, namely a black hole geometry supported by an additional effective source rather than by ordinary vacuum alone \cite{Contreras:2021yxe}. Consequently, the hairy corrections encoded in \(\tilde m(r)\) should be understood as a model-independent parametrization of extra matter-energy content or an additional gravitational sector surrounding the black hole. This interpretation is useful for our purposes, because it allows one to study the imprint of hair on particle dynamics without committing to a unique microscopic origin of the extra source.

%In the following sections, we shall analyze the motion of spinning test particles in the equatorial plane of the background (\ref{metric_rot_hairy}). Our main goal is to determine how the combined action of rotation and hair modifies the effective potential, the structure of stable and unstable circular orbits, the ISCO, and finally the Lyapunov exponent. 

In the following sections, we study generic spinning particle motion in
the rotating hairy background~\eqref{metric_rot_hairy}. We do not restrict the
Lyapunov analysis to equatorial circular orbits. Instead, we integrate
the full MPD system in phase-space and diagnose the separation of nearby trajectories. This allows us to examine how the combined effect of the particle spin and the hairy deformation modifies the global dynamical behavior of the orbit.

\section{Spinning test particles in the hairy black hole spacetime}
\label{sec:mpd}

We describe the motion of the spinning test particle by the
Mathisson--Papapetrou--Dixon (MPD) equations. In the pole--dipole
approximation, the internal structure of the body is represented by its
mass monopole and spin dipole, while higher multipole moments are
neglected. The equations of motion are
\begin{equation}
	\frac{D p^\mu}{d\tau}
	=
	-\frac{1}{2}
	R^\mu{}_{\nu\alpha\beta}
	v^\nu S^{\alpha\beta},
	\label{eq:mpd_force_tensor}
\end{equation}
and
\begin{equation}
	\frac{D S^{\mu\nu}}{d\tau}
	=
	p^\mu v^\nu-p^\nu v^\mu .
	\label{eq:mpd_spin_tensor}
\end{equation}
Here $p^\mu$ is the four-momentum of the particle,
$S^{\mu\nu}$ is the antisymmetric spin tensor, $v^\mu=dx^\mu/d\tau$
is the tangent vector to the representative worldline, and
$R^\mu{}_{\nu\alpha\beta}$ is the Riemann tensor of the rotating hairy
black hole spacetime.

The MPD equations must be supplemented by a spin supplementary condition
in order to select a unique representative center-of-mass worldline. We
use the Tulczyjew condition
\begin{equation}
	p_\mu S^{\mu\nu}=0 .
	\label{eq:tulczyjew_condition}
\end{equation}
The momentum normalization is chosen as
\begin{equation}
	p^\mu p_\mu=-\mu^2 ,
	\label{eq:momentum_norm_dimensional}
\end{equation}
where $\mu$ is the mass of the test particle.

The physical spin magnitude is defined by
\begin{equation}
	\frac{1}{2}S_{\mu\nu}S^{\mu\nu}=S_{\rm phys}^2 .
	\label{eq:spin_tensor_norm}
\end{equation}
Throughout the numerical analysis we use the dimensionless spin
parameter
\begin{equation}
	S \equiv \frac{S_{\rm phys}}{\mu M}.
	\label{eq:dimensionless_spin}
\end{equation}
The symbol $S$ will always denote this non-negative spin magnitude.
The spin orientation is specified independently by the initial spin
components in a local orthonormal frame. Thus, unlike in an equatorial
circular-orbit reduction, no signed spin variable is introduced in the
main analysis.

It is useful to work with momenta scaled by $\mu$ and lengths scaled by
$M$. In these units the momentum normalization becomes
\begin{equation}
	p^\mu p_\mu=-1,
	\label{eq:momentum_norm}
\end{equation}
and the dimensionless spin magnitude satisfies
\begin{equation}
	S^\mu S_\mu=S^2 ,
	\label{eq:spin_one_form_norm}
\end{equation}
where $S_\mu$ is the spin one-form introduced below.

\subsection{Spin-one-form formulation}

For the numerical evolution we use the spin-one-form formulation of the
MPD equations. The spin one-form is defined by
\begin{equation}
	S_\mu
	=
	\frac{1}{2}
	\epsilon_{\mu\nu\alpha\beta}
	p^\nu S^{\alpha\beta},
	\label{eq:spin_one_form_definition}
\end{equation}
where we have used the normalized momentum, $p^\mu p_\mu=-1$. The
inverse relation is
\begin{equation}
	S^{\mu\nu}
	=
	\epsilon^{\mu\nu\alpha\beta}
	p_\alpha S_\beta .
	\label{eq:spin_tensor_from_one_form}
\end{equation}
The Tulczyjew condition is then equivalent to
\begin{equation}
	p^\mu S_\mu=0 .
	\label{eq:ssc_one_form}
\end{equation}

In terms of the phase-space variables
\begin{equation}
	y=
	\left(
	x^\mu,p_\mu,S_\mu
	\right)
	=
	\left(
	t,r,\theta,\phi,
	p_t,p_r,p_\theta,p_\phi,
	S_t,S_r,S_\theta,S_\phi
	\right),
	\label{eq:phase_space_vector}
\end{equation}
the evolution equations used in the numerical integration are
\begin{equation}
	\dot{x}^{\mu}=v^\mu ,
	\label{eq:eom_x}
\end{equation}
\begin{equation}
	\dot{p}_{\mu}
	=
	\Gamma^{\alpha}{}_{\beta\mu}p_{\alpha}v^{\beta}
	-
	R^{*\alpha\beta}{}_{\mu\nu}
	v^\nu p_\alpha S_\beta ,
	\label{eq:eom_p}
\end{equation}
and
\begin{equation}
	\dot{S}_{\mu}
	=
	\Gamma^{\alpha}{}_{\beta\mu}S_{\alpha}v^{\beta}
	-
	p_{\mu}
	\left(
	R^{*\alpha}{}_{\beta\gamma\delta}
	S_\alpha v^\beta p^\gamma S^\delta
	\right).
	\label{eq:eom_S}
\end{equation}
Here an overdot denotes differentiation with respect to the proper time
$\tau$, $\Gamma^\alpha{}_{\beta\mu}$ are the Christoffel symbols of the
hairy metric, and $R^{*\alpha\beta}{}_{\mu\nu}$ denotes the right-dual
Riemann tensor. All geometric quantities are evaluated using the full
rotating hairy black hole metric.

The four-velocity is not parallel to the momentum when the spin is
nonzero. Instead, it is obtained from the MPD velocity--momentum relation
\begin{equation}
	v^\mu
	=
	N\left(p^\mu+w^\mu\right),
	\label{eq:velocity_momentum_relation}
\end{equation}
where
\begin{equation}
	w^\mu
	=
	-{}^{*}R^{*\mu\alpha\beta\gamma}
	S_\alpha p_\beta S_\gamma .
	\label{eq:w_correction}
\end{equation}
The normalization factor $N$ is fixed by
\begin{equation}
	v^\mu v_\mu=-1 .
	\label{eq:velocity_norm}
\end{equation}
Equations~\eqref{eq:eom_x}--\eqref{eq:velocity_norm} are the equations
integrated in the Lyapunov analysis below.

Hamiltonian and canonical descriptions of spinning particle motion
provide useful complementary approaches to the covariant MPD formulation
used here
\cite{Barausse:2009xi,Witzany:2018ahb,Witzany:2019nml,LukesGerakopoulos:2016hyy}.
Recent applications of spinning particle dynamics to Kerr scattering,
circular orbits, and comparisons with effective-one-body descriptions
can be found in Refs.~\cite{Gonzo:2024zxo,Piovano:2025kerr,
Albertini:2024mpd}.

\subsection{Conserved quantities}

The rotating hairy spacetime is stationary and axisymmetric. Therefore,
the Killing vectors $\partial_t$ and $\partial_\phi$ give two conserved
quantities for the spinning particle. For a general Killing vector
$\xi^\mu$, the corresponding conserved quantity is
\begin{equation}
	C_\xi
	=
	\xi^\mu p_\mu
	-
	\frac{1}{2}
	\xi_{\mu;\nu}S^{\mu\nu}.
	\label{eq:killing_conserved}
\end{equation}
The conserved energy and axial angular momentum are therefore
\begin{equation}
	E
	=
	-p_t
	+
	\frac{1}{2}
	g_{t\mu,\nu}S^{\mu\nu},
	\label{eq:energy_general}
\end{equation}
and
\begin{equation}
	J_z
	=
	p_\phi
	-
	\frac{1}{2}
	g_{\phi\mu,\nu}S^{\mu\nu}.
	\label{eq:jz_general}
\end{equation}
These expressions are used both in the construction of initial data and
as diagnostics for monitoring the accuracy of the numerical evolution.

\section{Parameterizing initial conditions}
\label{sec:initial_conditions}

The MPD equations form a constrained dynamical system. Therefore,
physically meaningful initial data cannot be constructed by assigning all
components of $x^\mu$, $p_\mu$, and $S_\mu$ independently. Instead, the
initial state must satisfy the momentum normalization, the spin
normalization, the Tulczyjew condition, and the conserved values of
$E$ and $J_z$.

For a broad survey of trajectories, it is useful to label the initial
data by geometrically motivated orbital parameters. We therefore
parameterize the requested orbit by
\begin{equation}
	(r_p,e,\iota),
\end{equation}
where $r_p$ is the requested pericenter, $e$ is the requested
eccentricity, and $\iota$ is the requested inclination. The background is
specified by $(a,\alpha,\beta)$, while the particle is specified by the
dimensionless spin magnitude $S$ and by the initial spin-orientation
components
\begin{equation}
	\hat S_r,\qquad \hat S_z ,
\end{equation}
defined in a local orthonormal frame. In the numerical survey $S$ is
always non-negative. The orientation of the spin is carried by
$\hat S_r$ and $\hat S_z$, not by the sign of $S$.

\subsection{Apocenter and requested turning points}

Given the requested pericenter and eccentricity, we define the corresponding apocenter as
\begin{equation}
 r_a=\left(\frac{1+e}{1-e}\right)r_p.
\end{equation}
For the seed orbit we introduce three auxiliary quantities
$(E_{\rm seed},J_{z,\rm seed},\mathcal{Q}_{\rm seed})$, where
$\mathcal{Q}_{\rm seed}$ is used only as an inclination-labeling
parameter. These are determined from the two requested radial turning
conditions
\begin{eqnarray}
    &&\mathcal{R}_{\rm h}(r_p;E_{\rm seed},J_{z,\rm seed},\mathcal{Q}_{\rm seed})=0,\\
    &&\mathcal{R}_{\rm h}(r_a;E_{\rm seed},J_{z,\rm seed},\mathcal{Q}_{\rm seed})=0,
\end{eqnarray}
together with the inclination condition \cite{Hartl:2003survey}
\begin{equation}
    \mathcal{Q}_{\rm seed}=J_{z,\rm seed}^2\tan^2\iota .
\end{equation}
Here $\mathcal{R}_{\rm h}$ denotes the Kerr-like radial seed function
with the Kerr $\Delta$ replaced by the hairy-background $\Delta(r)$.
The quantity $\mathcal{Q}_{\rm seed}$ should not be interpreted as an
exact Carter constant of the full hairy MPD system; it is only a seed
parameter used to generate the requested inclination
\begin{equation}
\begin{aligned}
\mathcal{R}_{\rm h}(r)
&=
\left[
E_{\rm seed}(r^2+a^2)-aJ_{z,\rm seed}
\right]^2  \\
&\quad
-\Delta(r)
\left[
r^2+
\left(J_{z,\rm seed}-aE_{\rm seed}\right)^2
+\mathcal{Q}_{\rm seed}
\right].
\end{aligned}
\label{eq:hairy_seed_radial_function}
\end{equation}
This radial function is used only to construct the requested seed orbit;
the subsequent evolution is performed with the full MPD equations.

This construction should be understood as an orbital-labeling strategy
rather than as an assertion that $(r_p,e,\iota)$ remain exact invariants
of the full spinning particle motion. A similar philosophy was used in
earlier Kerr MPD surveys, where geodesic orbital parameters were used to
construct initial data for spinning particle trajectories
\cite{Hartl:2003survey}. In the present work, the same idea is adapted
to the rotating hairy background. The important difference is that the
seed parameters are later compared with the empirical orbital parameters
extracted from the evolved MPD trajectory.
%The pair $(r_p,r_a)$ plays the role of the requested radial turning points. In the Kerr case one can map $(r_p,e,\iota)$ exactly to a set of conserved quantities through the separable geodesic equations. In the present hairy background the same interpretation is no longer exact, so the parameters $(r_p,e,\iota)$ should be understood as requested labels for the seed orbit rather than exact invariants of the final spinning motion. Nevertheless, they remain the most convenient and physically transparent coordinates in which to organize the survey.

\subsection{Seed orbit and starting point}

The first step is to determine a nonspinning seed orbit in the hairy background that reproduces the requested turning points. Operationally, the seed constants of motion are obtained by imposing the turning-point conditions at $r=r_p$ and $r=r_a$. The requested inclination is incorporated through the initial polar data so that the seed orbit crosses the equatorial plane with the desired tilt. The resulting seed orbit plays the same organizing role as the geodesic seed in the vacuum Kerr problem, but all geometric quantities are now evaluated in the hairy spacetime.

Once the seed orbit has been fixed, we choose the initial point of the spinning trajectory at an equatorial crossing,
\begin{equation}
 t_0=0, \qquad \phi_0=0, \qquad \theta_0=\frac{\pi}{2}.
\end{equation}
For the radial position we adopt the midpoint prescription
\begin{equation}
 r_0=\frac{r_p+r_a}{2},
\end{equation}
which is simple and robust for the range of eccentricities studied numerically. The initial radial momentum is then taken from the seed orbit at $r=r_0$.

\subsection{Construction of the full spinning particle state}

At this stage the initial conditions are only partially specified. To obtain a valid MPD state, one must solve the normalization conditions, the spin supplementary condition, and the exact conserved-quantity relations simultaneously. The two chosen spin components $\hat S_r$ and $\hat S_z$ are first converted from the local orthonormal frame into coordinate-basis covariant components at the starting point. The momentum normalization determines one remaining momentum component, while the spin normalization determines one missing spin component. The remaining unknowns are then obtained from the coupled nonlinear system formed by
\begin{enumerate}
 \item the spin supplementary condition,
 \item the exact energy relation,
 \item the exact axial angular-momentum relation.
\end{enumerate}
In practice this system is solved by a Newton--Raphson iteration. Every accepted initial condition is verified a posteriori to satisfy all constraints to numerical precision.

The final initial state may therefore be written as
\begin{equation}
 y_0=\left(t_0,r_0,\theta_0,\phi_0,p_t,p_r,p_\theta,p_\phi,S_t,S_r,S_\theta,S_\phi\right),
\end{equation}
where the momentum and spin components are not chosen independently but are determined self-consistently from the requested orbital data and the MPD constraints.

\subsection{Requested and empirical orbital parameters}

The requested orbital parameters do not, in general, coincide with the empirical parameters of the evolved spinning trajectory. Spin--curvature coupling alters the motion away from the seed orbit, and in the present problem the geometric hair further changes the strong field structure in which the orbit evolves. We therefore distinguish throughout between requested and empirical orbital parameters.

The empirical pericenter is extracted directly from the numerical trajectory by recording the minimum radius reached over a sufficiently long integration. Likewise, the empirical inclination is estimated from the maximal polar excursion away from the equatorial plane. In the weak spin and weak-hair limits, the empirical parameters reduce smoothly to the requested ones. In the strongly relativistic regime, however, the differences can be significant. This distinction is therefore essential when presenting parameter-space maps and interpreting the physical meaning of the survey.
\begin{comment}
\section{Lyapunov exponents and chaos detector}
\label{sec:lya}

To diagnose the onset of chaos we use the largest Lyapunov exponent. Let $y(\tau)$ be a reference trajectory in the twelve-dimensional phase-space and let $y'(\tau)$ be a nearby trajectory launched from a constraint preserving displaced initial condition. The deviation vector is
\begin{equation}
 \delta y(\tau)=y'(\tau)-y(\tau).
\end{equation}
If the motion is chaotic, the effective separation between the two trajectories grows exponentially,
\begin{equation}
 r_e(\tau)=\frac{\|\delta y(\tau)\|}{\|\delta y_0\|}\sim e^{\lambda \tau},
\end{equation}
so that the principal Lyapunov exponent is estimated from the slope of $\log r_e(\tau)$.
\end{comment}

%\section{Lyapunov exponents and chaos detector}
%\label{sec:lyapunov_method}
\section{Finite-Time Lyapunov Analysis}
\label{sec:lyapunov_method}
To diagnose chaos we evolve two nearby, constraint-preserving solutions
of the full MPD system. Let $y(\tau)$ be a reference solution of
Eqs.~\eqref{eq:eom_x}--\eqref{eq:eom_S}, and let $y'(\tau)$ be a nearby
solution launched from a slightly displaced initial condition satisfying
the same physical constraints. The deviation vector is
\begin{equation}
	\delta y(\tau)=y'(\tau)-y(\tau).
	\label{eq:deviation_vector}
\end{equation}
If the motion is chaotic, the effective separation grows approximately
as
\begin{equation}
	r_e(\tau)
	=
	\frac{\|\delta y(\tau)\|}{\|\delta y(0)\|}
	\sim e^{\lambda \tau},
	\label{eq:deviation_growth}
\end{equation}
and the largest Lyapunov exponent is estimated from the growth of
$\log r_e(\tau)$.

\subsection{Deviation-vector method}

For a broad survey of parameter space, a full Jacobian evolution is computationally costly. We therefore adopt the unrescaled deviation-vector approach, in which only the two nearby trajectories are evolved. This method is substantially faster and is adequate for mapping the broad chaotic structure of the system. At regular sampling intervals we record $\log r_e(\tau)$ and determine the exponent by a least-squares fit over the linear-growth regime.

%The use of an unrescaled deviation vector, together with a pre-saturation fitting window, follows the spirit of earlier numerical studies of MPD chaos in Kerr spacetime~\cite{Hartl:2003survey}. The main difference here is that both trajectories are evolved in the full rotating hairy geometry and the deviation is measured using the projected phase-space norm defined below.

The use of a deviation vector method and a pre saturation fitting window
is closely related to earlier numerical studies of spinning particle
chaos in black hole spacetimes \cite{Suzuki:1997chaos,Hartl:2002ig,Hartl:2003survey,Han:2010tp,Zelenka:2019nyp}. 
%\cite{Hartl:2002ig,Hartl:2003survey,Han:2010tp}.
In those works, the growth of the separation between nearby MPD
trajectories was used as a practical diagnostic of chaotic behavior.
Here we adopt the same general philosophy, but evolve both trajectories
in the rotating hairy black hole background and measure their separation
with the projected phase-space norm defined below.

The method has one important limitation: eventually the deviation saturates, because the two trajectories separate so far that the difference vector no longer samples the local instability of the flow. For that reason the exponent is not read off from arbitrarily late times, but from the pre-saturation growth regime.

\subsection{Projected phase-space norm}
\label{subsec:projected_norm}

The norm $\|\delta y\|$ is not taken to be a naive Euclidean norm in
the coordinate variables. Such a coordinate norm would depend strongly
on the choice of coordinates and would be especially inconvenient in a
rotating spacetime, where the time and azimuthal directions are mixed by
frame dragging. Instead, we measure the separation between nearby
trajectories in the local spatial frame of a zero-angular-momentum
observer (ZAMO). Projected norms of this type were used in earlier analyses of spinning particle chaos in Kerr spacetime in order to reduce coordinate artifacts in the measurement of phase-space separation
\cite{Hartl:2003survey}.

%Projected norms of this type have been used in previous analyses of spinning particle chaos in Kerr spacetime in order to avoid interpreting coordinate artifacts as physical trajectory separation \cite{Karas:1992,Hartl:2003survey}.

For a stationary and axisymmetric metric, the ZAMO angular velocity is
\begin{equation}
    \omega
    =
    -\frac{g_{t\phi}}{g_{\phi\phi}} .
    \label{eq:zamo_omega}
\end{equation}
The corresponding ZAMO four-velocity can be written as
\begin{equation}
u^\mu_{\rm Z}
=
\frac{1}{\mathcal{N}_{\rm Z}}
(1,0,0,\omega),
\end{equation}
%\begin{equation}
 %   u^\mu_{\rm Z}
  %  =
   % \frac{1}{\mathcal{N}}
    %\left(
    %1,0,0,\omega
    %\right),
    %\label{eq:zamo_velocity}
%\end{equation}
where the normalization factor $\mathcal{N_{\rm Z}}$ is fixed by
$u^\mu_{\rm Z}u^{\rm Z}_\mu=-1$. Equivalently,
\begin{equation}
    \mathcal{N_{\rm Z}}
    =
    \sqrt{
    -g_{tt}
    -2\omega g_{t\phi}
    -\omega^2 g_{\phi\phi}
    } .
    \label{eq:zamo_lapse}
\end{equation}
The spatial projector orthogonal to the ZAMO worldline is then
\begin{equation}
    h_{\mu\nu}
    =
    g_{\mu\nu}
    +
    u^{\rm Z}_{\mu}u^{\rm Z}_{\nu}.
    \label{eq:zamo_projector_cov}
\end{equation}
For the coordinate displacement between two nearby trajectories,
$\delta x^\mu=x'^\mu-x^\mu$, the projected spatial separation is
defined by
\begin{equation}
    \|\delta x\|_{\rm Z}^{2}
    =
    h_{\mu\nu}\delta x^\mu\delta x^\nu .
    \label{eq:zamo_position_norm}
\end{equation}

The momentum and spin differences are covectors in our phase-space
description. Therefore we use the inverse spatial projector
\begin{equation}
    h^{\mu\nu}
    =
    g^{\mu\nu}
    +
    u^\mu_{\rm Z}u^\nu_{\rm Z}
    \label{eq:zamo_projector_contra}
\end{equation}
to define
\begin{equation}
    \|\delta p\|_{\rm Z}^{2}
    =
    h^{\mu\nu}\delta p_\mu\delta p_\nu ,
    \qquad
    \|\delta S\|_{\rm Z}^{2}
    =
    h^{\mu\nu}\delta S_\mu\delta S_\nu .
    \label{eq:zamo_momentum_spin_norms}
\end{equation}
The total projected phase-space separation is then taken to be
\begin{equation}
    \|\delta y\|_{\rm Z}
    =
    \left[
    \|\delta x\|_{\rm Z}^{2}
    +
    \|\delta p\|_{\rm Z}^{2}
    +
    \|\delta S\|_{\rm Z}^{2}
    \right]^{1/2}.
    \label{eq:zamo_total_norm}
\end{equation}
All quantities are evaluated along the reference trajectory. Since the
variables used in the numerical calculation are expressed in units of
$M$ and $\mu$, the three contributions in
Eq.~\eqref{eq:zamo_total_norm} are dimensionless in the code. The
deviation factor used in the Lyapunov analysis is therefore
\begin{equation}
    r_e(\tau)
    =
    \frac{\|\delta y(\tau)\|_{\rm Z}}
    {\|\delta y(0)\|_{\rm Z}} .
    \label{eq:zamo_deviation_factor}
\end{equation}
This projected norm gives a local physical measure of the phase-space
separation and avoids interpreting coordinate artifacts as exponential
instability.

%\subsection{Projected phase-space norm}

%The norm $\|\delta y\|$ is not taken to be a naive Euclidean norm in the coordinate variables. Instead, we project the position, momentum, and spin differences onto the local spatial hypersurface associated with zero-angular-momentum observers. The projected quantities are then combined through a Euclidean norm. This provides a physically meaningful local measure of phase-space separation in a rotating spacetime and avoids the coordinate ambiguities that would arise from using raw coordinate differences.

\subsection{Constrained deviations and false positives}

Because the MPD system is constrained, the nearby initial state must lie on the same physical constraint surface as the reference state. The initial deviation is therefore constructed so that the displaced trajectory satisfies the momentum normalization, the spin normalization, the spin supplementary condition, and the conserved quantities to the same numerical tolerance as the reference trajectory. This is crucial for a meaningful Lyapunov analysis.

In addition, highly relativistic regular trajectories can sometimes mimic chaotic growth for finite intervals. This occurs particularly for orbits with large precession or zoom--whirl behavior, where the phase-space separation can undergo large bursts without developing a persistent positive asymptotic slope. To avoid false positives, we do not classify an orbit as chaotic merely because the deviation becomes temporarily large. Instead, we require repeated large-separation points before identifying a positive chaotic signal, and we supplement the survey with deeper integrations of representative borderline trajectories.

\subsection{Numerical implementation}
\label{subsec:numerical_implementation}

The equations of motion are integrated in proper time. At every step the
full hairy geometry is reconstructed, including the metric, inverse
metric, Christoffel symbols, curvature tensors, dual curvature tensors,
and the MPD four-velocity obtained from
Eq.~\eqref{eq:velocity_momentum_relation}. The reference trajectory and
the nearby trajectory are evolved with the same integration scheme and
the same step size.

For the production runs we use a fixed-step fourth-order Runge--Kutta
integrator with step size
\begin{equation}
    \Delta\tau = 0.05M.
\end{equation}
The step size is chosen small enough that the conserved
quantities and constraints remain stable over the full integration time.
In the scans reported below, the typical integration time is
\begin{equation}
    \tau_{\rm max}=10^5 M,
\end{equation}
and the initial phase-space separation is
\begin{equation}
    \epsilon_0 = 3\times10^{-7}.
\end{equation}
The deviation vector is sampled every $N_{\rm samp}=2000$ integration steps. This corresponds to a sampling interval \(\Delta\tau_{\rm samp}=N_{\rm samp}\Delta\tau=100M\).  For selected representative trajectories we repeated the integration
with smaller values of $\Delta\tau$ to check that the qualitative
finite-time Lyapunov behavior is unchanged.

During the evolution we monitor the following constraint quantities:
\begin{equation}
    C_p = \left|p^\mu p_\mu+1\right|,
    \qquad
    C_S = \left|S^\mu S_\mu-S^2\right|,
\end{equation}
\begin{equation}
    C_{\rm SSC} = \left|p^\mu S_\mu\right|,
\end{equation}
and the relative conservation errors
\begin{equation}
    C_E =
    \left|\frac{E(\tau)-E(0)}{E(0)}\right|,
    \qquad
    C_J =
    \left|\frac{J_z(\tau)-J_z(0)}{J_z(0)}\right|.
\end{equation}
A trajectory is accepted only when these quantities remain below the
chosen numerical tolerance throughout the part of the evolution used in
the Lyapunov fit. In the present calculations this tolerance is taken to
be
\begin{equation}
    C_p,\ C_S,\ C_{\rm SSC},\ C_E,\ C_J < \varepsilon_{\rm tol},
\end{equation}
where $\varepsilon_{\rm tol}\sim 10^{-8}-10^{-6},$ is fixed by the accuracy of the numerical
run. Trajectories that violate this condition, or that plunge into the
near-horizon region, are excluded from the statistics of bound chaotic
motion.

The largest Lyapunov exponent is estimated from the growth of
$\log r_e(\tau)$ in the pre-saturation regime. We first discard an
initial transient interval,
\begin{equation}
    \tau < \tau_{\rm min},
\end{equation}
because the early-time separation can depend on the direction of the
initial deviation vector. We then fit $\log r_e(\tau)$ over the interval
in which it exhibits approximately linear growth. The fit is stopped
when the deviation reaches the saturation threshold,
\begin{equation}
    \log r_e(\tau) \geq \log r_{e,\rm sat}.
\end{equation}
In practice, $r_{e,\rm sat}$ corresponds to the point at which the two
trajectories are no longer close enough for the deviation vector to
probe the local instability of the reference orbit. The finite-time
Lyapunov exponent reported in the figures is therefore obtained from
\begin{equation}
    \lambda_{\max}
    =
    \frac{d}{d\tau}\log r_e(\tau)
\end{equation}
as determined by a least-squares fit over the accepted pre-saturation
window.

This procedure avoids two common numerical ambiguities. First, it
prevents the late-time saturated separation from being interpreted as
continued exponential growth. Second, it reduces false positives from
regular but highly relativistic trajectories that show temporary bursts
in $\log r_e(\tau)$ due to precession or zoom--whirl-like motion.

As consistency checks, we verified that in the limit $S=0$ the MPD
trajectory reduces to geodesic-like motion in the same background, while
for $\alpha=0$ the metric reduces to Kerr and the qualitative spin
dependence agrees with the known behavior of spinning particle dynamics
in Kerr spacetime. We also monitored the conservation of $E$ and $J_z$
throughout the evolution and discarded trajectories for which the
constraint violations exceeded the prescribed tolerance.

\begin{table}[t]
\centering
\begin{tabular}{c|c}
\hline\hline
Quantity & Value/range \\
\hline
black hole spin & $a=0.9$ \\
Eccentricity & $e=0.5$ \\
Requested pericenter scan & $2M\leq r_p\leq 4M$ \\
Requested inclination scan & $5^\circ\leq \iota\leq 85^\circ$ \\
Spin values & $S=0,0.01,0.05,0.1,0.5,0.9,1.0$ \\
Hair values in trajectory plots & $\beta=0.2,0.8,1.5$ \\
Hair scan in $(S,\beta)$ map & $0\leq \beta \leq 1.9$ \\
Initial separation & $\epsilon_0=3\times 10^{-7}$ \\
Maximum integration time & $\tau_{\rm max}=10^5M$ \\
Sampling interval & $N_{\rm samp}=2000$ steps \\
Integrator step size & $\Delta\tau=0.05M$ \\
\hline\hline
\end{tabular}
\caption{
Representative numerical parameters used in the Lyapunov survey.
The large values $S=0.9$ and $S=1.0$ are used as dynamical probes of
strong spin--curvature coupling rather than as direct astrophysical
EMRI spin values.
}
\label{tab:numerical_setup}
\end{table}

In the time domain plots, not all trajectories necessarily extend to the
same final integration time. This is a consequence of the stopping
criteria used in the Lyapunov analysis. If the deviation factor reaches
the prescribed saturation threshold, the Lyapunov fit is terminated and
no further data from that trajectory are used in the exponent estimate.
Similarly, trajectories that plunge into the near-horizon cutoff region
or violate the constraint tolerances are stopped and classified
separately. Therefore, a shorter curve in the $\lambda(\tau)$ or
$\log r_e(\tau)$ plots should not be interpreted as missing data. It
means that the corresponding trajectory has reached one of the
predefined termination conditions.

%\subsection{Numerical implementation}

%The equations of motion are integrated in proper time using a high-accuracy numerical integrator. At every step the full hairy geometry is reconstructed, including the metric, inverse metric, Christoffel symbols, curvature tensors, and corrected four-velocity. The numerical evolution is monitored by tracking the momentum normalization, the spin normalization, the spin supplementary condition, and the conservation of $E$ and $J_z$. Trajectories that plunge into the near-horizon region are classified separately and removed from the statistics of bound chaotic motion.

%\section{Results and discussion}\label{sec:results}

%The results are organised here in a way that first exploits the intuition gained from the effective-potential and ISCO analysis, and then moves to the global phase-space survey.

%\subsection{Requested-parameter and empirical-parameter maps}
\section{Results and discussion}
\label{sec:results}

We now present the numerical results obtained by integrating the full
MPD system~\eqref{eq:eom_x}--\eqref{eq:eom_S} in the rotating hairy
black hole background. The results are organized around three related
questions: how the requested initial-data grid is mapped into empirical
orbital parameters, how the particle spin controls the growth of nearby
trajectories, and how the spin magnitude and hair parameter jointly
shape the largest Lyapunov exponent.

%\begin{table}[t]
%\centering
%\begin{tabular}{c|c}
%Quantity & Value/range \\
%\hline
%black hole spin & $a=0.9$ \\
%Eccentricity & $e=0.5$ \\
%Requested pericenter scan & $2M\leq r_p\leq 4M$ \\
%Requested inclination scan & $5^\circ\leq \iota\leq 85^\circ$ \\
%Spin values & $S=0,0.01,0.05,0.1,0.5,0.9,1.0$ \\
%Hair values & $\beta=0.2,0.8,1.5$ \\
%Initial separation & $\epsilon_0=3\times10^{-7}$ \\
%Integration time & $\tau_{\rm max}=10^{5}$ \\
%\end{tabular}
%\caption{Representative numerical parameters used in the Lyapunov survey.}
%\label{tab:numerical_setup}
%\end{table}

\subsection{Requested-parameter and empirical-parameter maps}
\label{subsec:req_emp_maps}

We first examine the distribution of the largest finite-time Lyapunov
exponent in the $(r_p,\iota)$ plane. For each point in the numerical grid,
we construct a constraint-preserving MPD initial condition from the requested
orbital data, evolve the corresponding spinning-particle trajectory in the
rotating hairy background, and compute the principal Lyapunov exponent using
the deviation-vector method described above. The result is displayed in two
complementary ways. In the requested-parameter maps the horizontal and vertical
axes are the input parameters $(r_{p,{\rm req}},\iota_{\rm req})$ used in the
initial-condition construction. In the empirical-parameter maps the same
trajectory is instead placed at the measured values
\begin{equation}
	r_{p,{\rm emp}}=\min_\tau r(\tau),
	\qquad
	\iota_{\rm emp}=\max_\tau \left|\frac{\pi}{2}-\theta(\tau)\right| .
\end{equation}
The color scale is identical in the two panels of each figure and represents
the same value of $\lambda_{\max}$. Thus, the requested map shows where the
large finite-time Lyapunov signal appears in the initial-data grid, while the
empirical map shows where the corresponding orbit actually resides after the
full MPD evolution. This distinction is important because the spin--curvature
force and the hairy deformation both shift the radial turning points and the
polar excursion away from the seed orbit.

\begin{figure*}[t]
	\centering
	\begin{subfigure}[b]{0.45\textwidth}
		\centering
		\includegraphics[width=\linewidth]{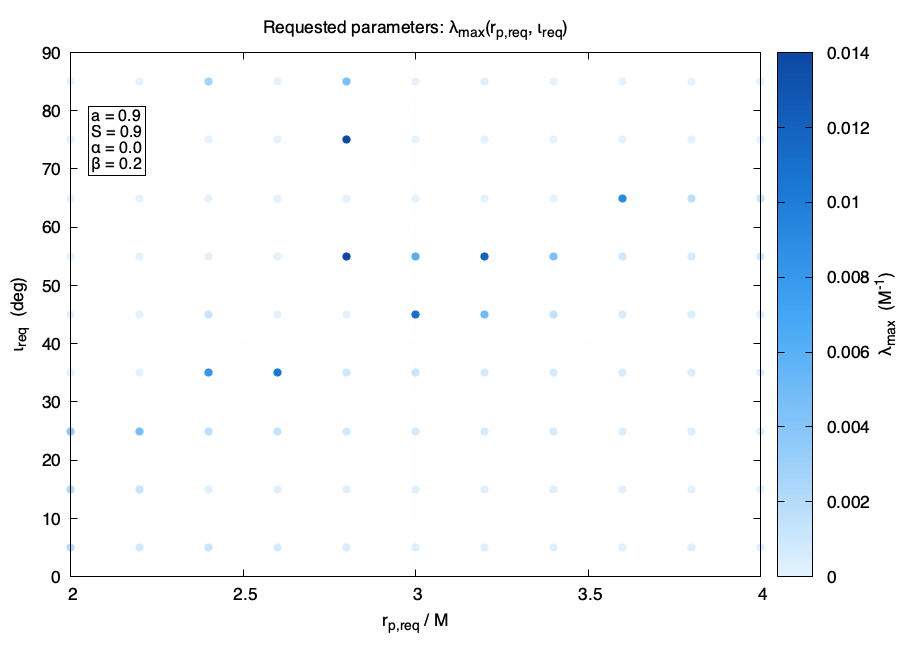}
		\caption{Requested parameters.}
		\label{fig:req_emp_kerr_req}
	\end{subfigure}
	\hfill
	\begin{subfigure}[b]{0.45\textwidth}
		\centering
		\includegraphics[width=\linewidth]{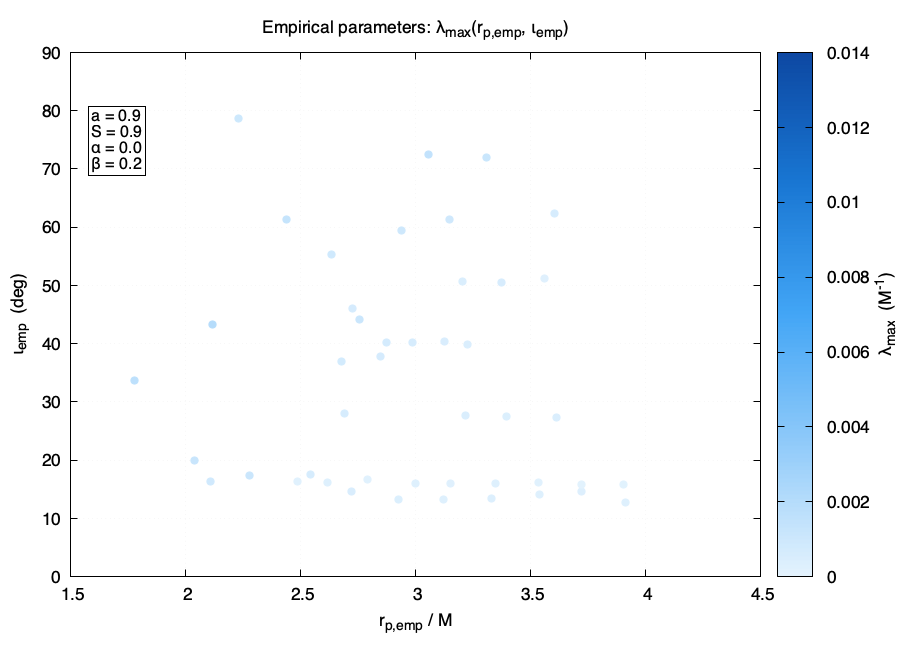}
		\caption{Empirical parameters.}
		\label{fig:req_emp_kerr_emp}
	\end{subfigure}
	\caption{Largest finite-time Lyapunov exponent in the $(r_p,\iota)$
		plane for $a=0.9$, $S=0.9$, $\alpha=0$, and $\beta=0.2$. Since the
		deformation term is proportional to $\alpha$, this case corresponds to
		the Kerr limit. Panel (a) shows the requested parameters used to generate
		the initial data, while panel (b) shows the same trajectories replotted
		using the empirical pericenter and inclination measured from the evolved
		orbit. The common color scale gives $\lambda_{\max}$ in units of
		$M^{-1}$.}
	\label{fig:req_emp_kerr}
\end{figure*}

Figure~\ref{fig:req_emp_kerr} provides the reference case in which the hairy
deformation is switched off. The requested map forms the expected rectangular
sampling grid in $(r_{p,{\rm req}},\iota_{\rm req})$. Most of the grid points
show weak or negligible finite-time exponential growth, while the larger values
of $\lambda_{\max}$ are confined to isolated regions. These larger values occur
mainly for selected moderate-to-large inclinations and for pericenters in the
strong-field part of the scanned interval. This already shows that the finite-
time instability is not controlled by one requested orbital label alone; rather,
it depends on the combined choice of pericenter, inclination, black-hole
rotation, and particle spin.

The empirical panel in Fig.~\ref{fig:req_emp_kerr}(b) shows that the same set of
trajectories is no longer arranged on a rectangular grid after the MPD
evolution. The points are redistributed over a wider region of empirical
parameter space, with several trajectories shifted toward smaller empirical
pericenters and with the inclination distributed over both low- and high-
inclination regions. This redistribution occurs even in the Kerr limit and is
therefore a consequence of spin--curvature coupling in the MPD dynamics. The
empirical map is thus essential for interpreting the physical location of the
large Lyapunov signal: the requested parameters label how the initial data were
generated, whereas the empirical parameters reveal the actual orbital region
sampled by the evolved trajectory.

\begin{figure*}[t]
	\centering
	\begin{subfigure}[b]{0.45\textwidth}
		\centering
		\includegraphics[width=\linewidth]{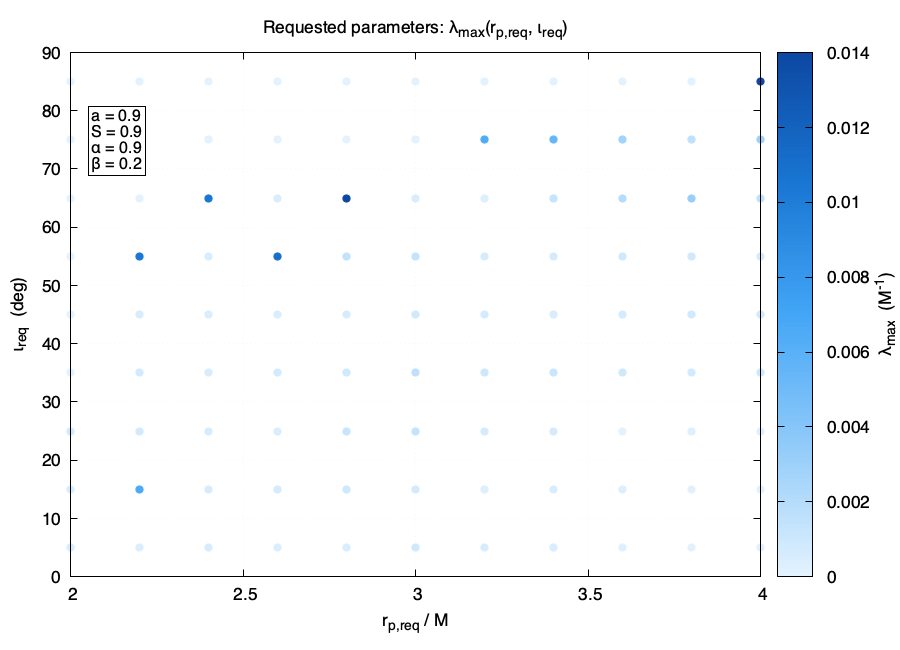}
		\caption{Requested parameters.}
		\label{fig:req_emp_hairy_b02_req}
	\end{subfigure}
	\hfill
	\begin{subfigure}[b]{0.45\textwidth}
		\centering
		\includegraphics[width=\linewidth]{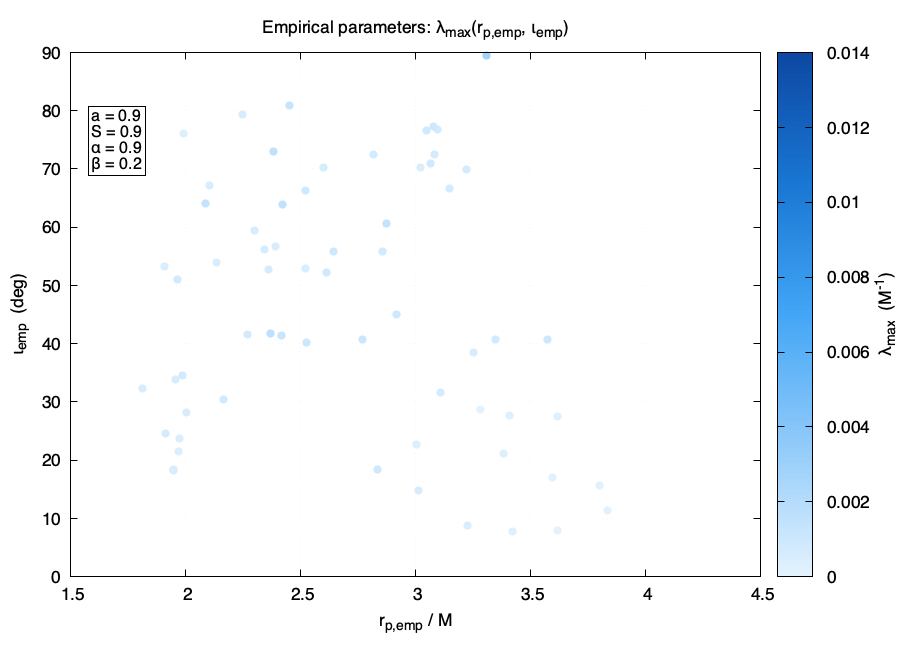}
		\caption{Empirical parameters.}
		\label{fig:req_emp_hairy_b02_emp}
	\end{subfigure}
	\caption{Largest finite-time Lyapunov exponent in the $(r_p,\iota)$
		plane for $a=0.9$, $S=0.9$, $\alpha=0.9$, and $\beta=0.2$. Compared with
		the Kerr limit map in Fig.~\ref{fig:req_emp_kerr}, the nonzero hairy
		deformation changes both the location of the larger Lyapunov points in the
		requested grid and the redistribution of the same trajectories in
		empirical parameter space. The common color scale gives $\lambda_{\max}$
		in units of $M^{-1}$.}
	\label{fig:req_emp_hairy_b02}
\end{figure*}

Figure~\ref{fig:req_emp_hairy_b02} shows the first genuinely hairy case, with
$\alpha=0.9$ and $\beta=0.2$. In the requested map, the finite-time Lyapunov
signal remains sparse, but the locations of the stronger points are shifted
relative to the Kerr reference case. The larger values now appear at several
moderate and high requested inclinations, including configurations near the
outer part of the requested radial range. This indicates that the hairy
correction does not simply rescale the Kerr pattern. Instead, it changes which
requested seed configurations evolve into trajectories with stronger sensitivity
to initial conditions.

The empirical map in Fig.~\ref{fig:req_emp_hairy_b02}(b) makes this effect more
transparent. The evolved trajectories populate a broad empirical region rather
than preserving the rectangular structure of the requested grid. In particular,
the empirical inclinations extend over a wide range, including high-inclination
trajectories close to the upper part of the plotted interval. The empirical
pericenters are also redistributed across the strong-field region. Therefore,
part of the effect of the hairy deformation is to alter the map between the
requested seed labels and the actual strong-field trajectory sampled by the
spinning particle. This is important because the Lyapunov exponent is controlled
by the geometry along the evolved orbit, not only by the labels used to generate
the initial condition.

\begin{figure*}[t]
	\centering
	\begin{subfigure}[b]{0.45\textwidth}
		\centering
		\includegraphics[width=\linewidth]{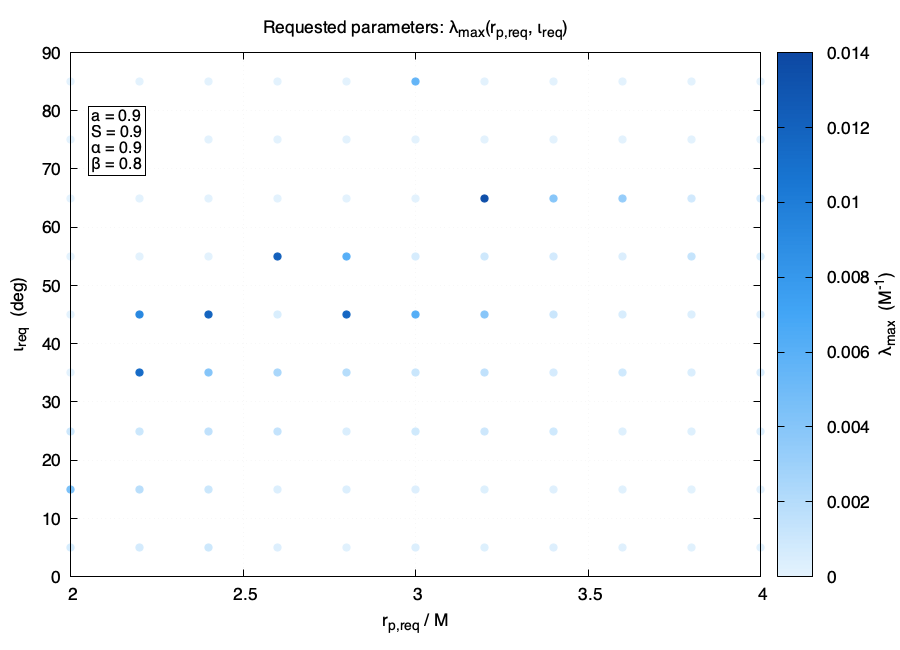}
		\caption{Requested parameters.}
		\label{fig:req_emp_hairy_b08_req}
	\end{subfigure}
	\hfill
	\begin{subfigure}[b]{0.45\textwidth}
		\centering
		\includegraphics[width=\linewidth]{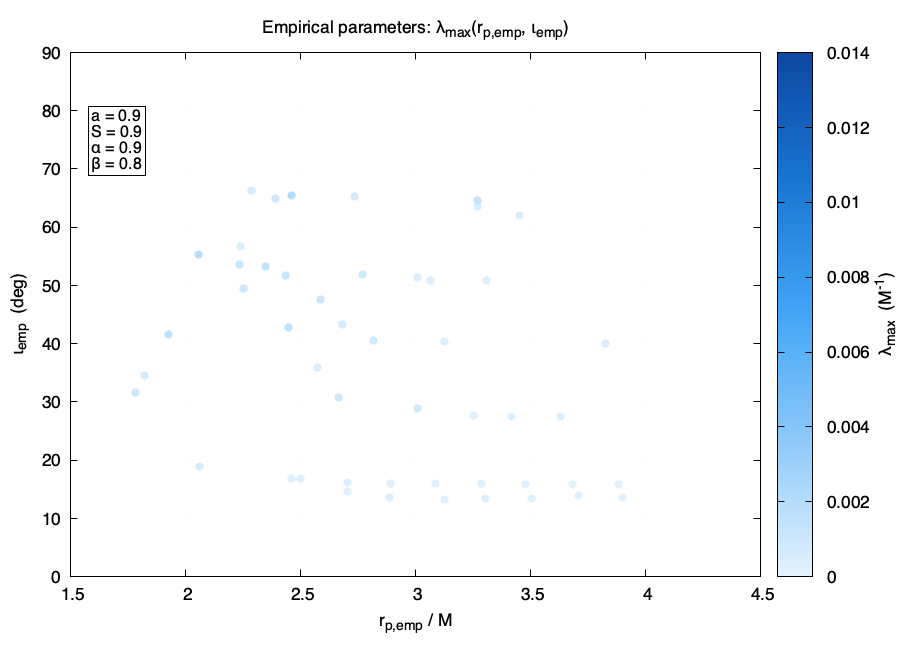}
		\caption{Empirical parameters.}
		\label{fig:req_emp_hairy_b08_emp}
	\end{subfigure}
	\caption{Largest finite-time Lyapunov exponent in the $(r_p,\iota)$
		plane for $a=0.9$, $S=0.9$, $\alpha=0.9$, and $\beta=0.8$. The change in
		$\beta$ modifies the distribution of the larger Lyapunov points in the
		requested grid and produces a different empirical redistribution of the
		evolved trajectories. The common color scale gives $\lambda_{\max}$ in
		units of $M^{-1}$.}
	\label{fig:req_emp_hairy_b08}
\end{figure*}

The effect of changing the hair parameter is displayed in
Fig.~\ref{fig:req_emp_hairy_b08}. For $\beta=0.8$, the requested map shows a
noticeable rearrangement of the larger Lyapunov points. The enhanced values are
not spread uniformly over the grid; instead, they occur at selected
low-to-intermediate and moderate-to-high inclinations. This confirms that
changing $\beta$ modifies the strong-field orbital response in a nontrivial way.
For the deformed mass function used here, $\beta$ enters through the radial
scale of the exponential deformation. Thus, varying $\beta$ changes the radial
localization of the hairy correction, rather than merely increasing a single
``hair strength'' parameter.

The corresponding empirical map shows that the $\beta=0.8$ trajectories are
again redistributed away from the requested grid. Compared with the $\beta=0.2$
case, the empirical points are less uniformly spread over the full inclination
range and show a clearer concentration in low- and moderate-inclination bands,
with additional points at higher empirical inclination. This demonstrates that
changing the radial scale of the deformation changes not only the value of
$\lambda_{\max}$ but also the requested-to-empirical map itself. In this sense,
the hair affects the finite-time instability indirectly, by changing the
strong-field region actually sampled by the MPD trajectory.

\begin{figure*}[t]
	\centering
	\begin{subfigure}[b]{0.45\textwidth}
		\centering
		\includegraphics[width=\linewidth]{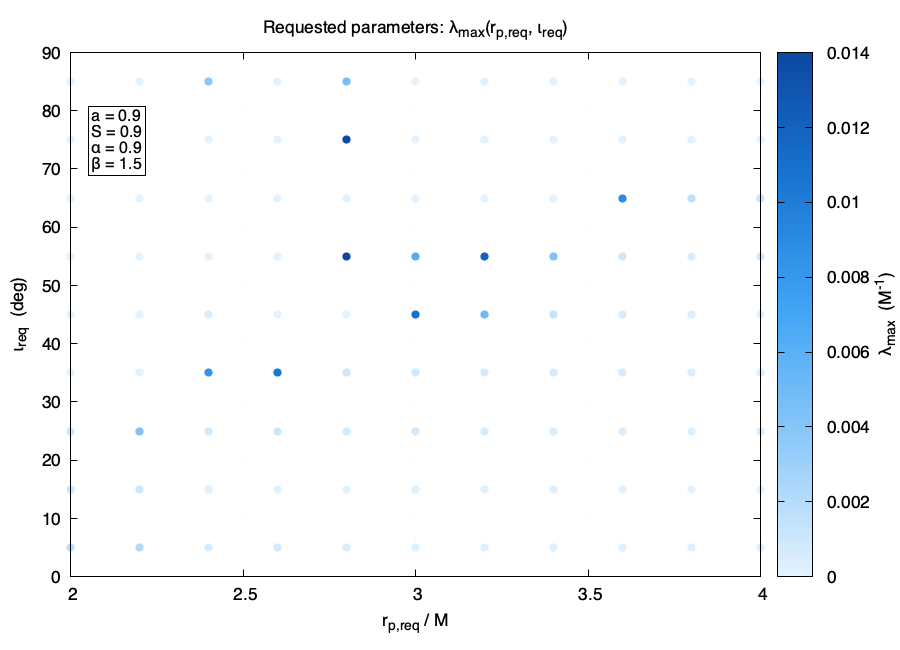}
		\caption{Requested parameters.}
		\label{fig:req_emp_hairy_b15_req}
	\end{subfigure}
	\hfill
	\begin{subfigure}[b]{0.45\textwidth}
		\centering
		\includegraphics[width=\linewidth]{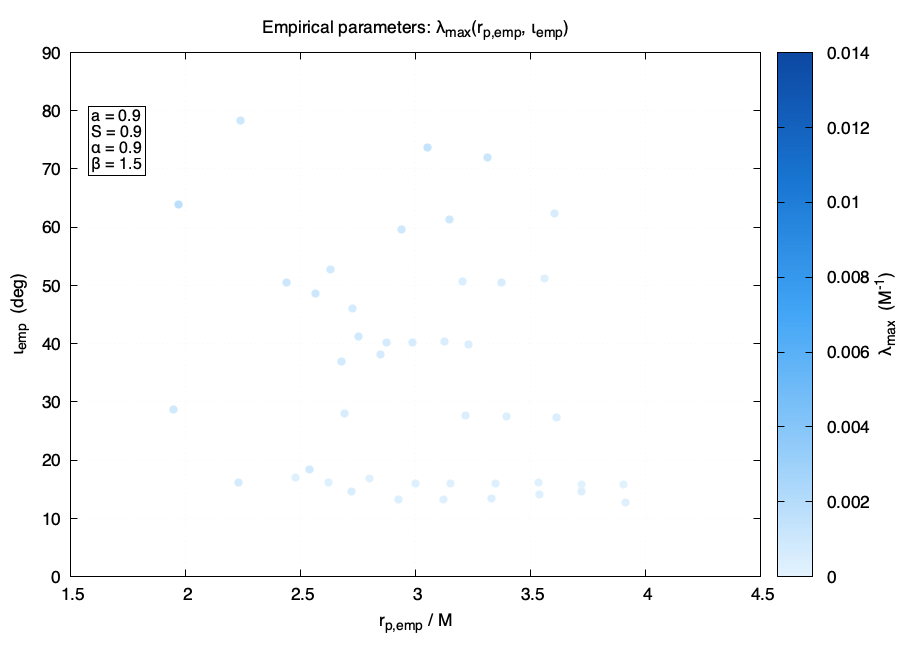}
		\caption{Empirical parameters.}
		\label{fig:req_emp_hairy_b15_emp}
	\end{subfigure}
	\caption{Largest finite-time Lyapunov exponent in the $(r_p,\iota)$
		plane for $a=0.9$, $S=0.9$, $\alpha=0.9$, and $\beta=1.5$. The larger
		value of $\beta$ changes the radial localization of the hairy deformation.
		The distribution of the stronger Lyapunov points and the empirical
		redistribution are therefore not a monotonic continuation of the
		$\beta=0.2$ and $\beta=0.8$ cases. The common color scale gives
		$\lambda_{\max}$ in units of $M^{-1}$.}
	\label{fig:req_emp_hairy_b15}
\end{figure*}

Figure~\ref{fig:req_emp_hairy_b15} shows the larger-$\beta$ case. The requested
map retains a sparse distribution of stronger Lyapunov points, but the pattern
is not simply an amplified version of the $\beta=0.8$ map. Instead, the
locations of the larger values partially return toward a Kerr-like sparse
structure, with prominent points appearing only at selected inclinations and
pericenters. The empirical map also becomes less broadly distributed than in the
$\beta=0.2$ case, with many trajectories occupying low- and moderate-inclination
regions. This behavior is consistent with the interpretation that increasing
$\beta$ changes the radial support of the exponential deformation and can move
the evolved orbit either toward or away from the most unstable part of the
strong-field phase space.

Taken together, Figs.~\ref{fig:req_emp_kerr}--\ref{fig:req_emp_hairy_b15} show
that the finite-time Lyapunov response is highly localized in the
$(r_p,\iota)$ plane. The strongest growth does not occur for every small
pericenter or every large inclination. Instead, it appears only when the
requested orbital geometry is mapped by the full MPD evolution into an empirical
trajectory that probes the appropriate strong-field curvature region. The four
pairs of maps therefore show that the requested--empirical distinction is not
merely a bookkeeping device; it is part of the physical mechanism by which the
hairy deformation reorganizes the phase-space region sampled by the spinning
trajectory. In the following subsections we use this distinction to analyze the
dependence on the particle spin and on the hair parameter more directly.

\subsection{Dependence on the particle spin}
\label{subsec:spin_dependence_running_lambda}

We next examine how the finite-time Lyapunov behavior changes when
the particle spin is varied while the orbital and background parameters
are kept fixed.  For each background we evolve the same representative
orbit for several values of the spin magnitude.  Before presenting the
spin scan, however, we emphasize the interpretation of the large-spin
cases.  The dimensionless quantity used here is
\begin{equation}
	S=\frac{S_{\rm phys}}{\mu M}.
\end{equation}
For a strict astrophysical extreme-mass-ratio system, a compact body
with dimensionless Kerr spin $\chi$ would typically have
$S_{\rm phys}\sim \chi\mu^2$, and therefore
\begin{equation}
	S\sim \chi\frac{\mu}{M}\ll 1 .
\end{equation}
More generally, the pole--dipole MPD approximation has a formal
validity condition: the spin length of the small body must remain much
smaller than the local curvature radius sampled by the orbit.  In
invariant form this requires
\begin{equation}
	\epsilon_{\rm spin}
	\equiv
	\frac{S_{\rm phys}}{\mu \mathcal{L}_{\rm curv}}
	\ll 1 ,
\end{equation}
where $\mathcal{L}_{\rm curv}$ denotes the characteristic local radius
of curvature.  In the strong-field region of a black hole one typically
has $\mathcal{L}_{\rm curv}\sim M$, so the condition reduces
approximately to
\begin{equation}
	S=\frac{S_{\rm phys}}{\mu M}\ll 1 .
\end{equation}
Thus, the large values $S=0.5,0.9,$ and $1.0$ used below should not be
interpreted as direct astrophysical EMRI spin values within the strict
pole--dipole regime.  Rather, they are used as controlled dynamical
probes of the nonlinear MPD phase space.  The physically perturbative
regime is represented by the small-spin cases, while the large-spin
cases reveal where strong spin--curvature coupling can drive enhanced
finite-time instability.  In this large-spin regime, higher multipole
and finite-size effects may become important, and the results should be
interpreted with this limitation in mind.

Recent work has also suggested that chaotic behavior may persist for
astrophysically realistic secondary spins in Schwarzschild spacetime
and may leave discernible gravitational-wave signatures
\cite{Yuan:2026realisticspin}.  This further motivates the study of how
spin-induced sensitivity to initial conditions is modified once the
background geometry is deformed away from Kerr.

We then consider
\begin{equation}
	S=1.0,\;0.9,\;0.5,\;0.1,\;0.05,\;0.01,\;0 ,
\end{equation}
and compute the running finite-time exponent
\begin{equation}
	\lambda(\tau)=\frac{\log r_e(\tau)}{\tau}.
\end{equation}
This quantity is not interpreted as the final asymptotic exponent at
every value of $\tau$; rather, it is used as a diagnostic for whether
the separation between nearby MPD trajectories develops sustained
positive growth before the deviation vector reaches the nonlinear
saturation regime.  For a regular trajectory, $\log r_e(\tau)$ grows at
most slowly and the running value $\lambda(\tau)$ decreases toward
zero.  For a trajectory showing strong finite-time exponential
sensitivity, $\log r_e(\tau)$ develops an approximately linear growth
interval and the corresponding running exponent remains comparatively
larger over the integration time.

\begin{figure*}[t]
	\centering
	\begin{minipage}{0.48\textwidth}
		\centering
		\includegraphics[width=\linewidth]{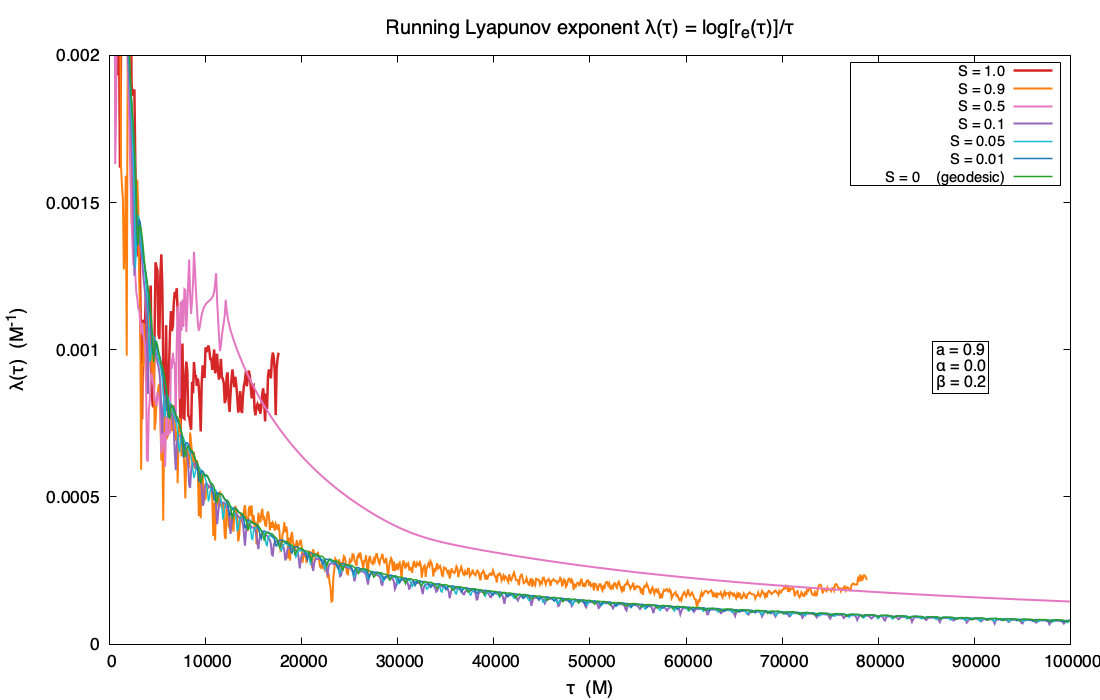}
		\vspace{1mm}
		\centerline{(a) Kerr limit: $a=0.9$, $\alpha=0$, $\beta=0.2$.}
	\end{minipage}
	\hfill
	\begin{minipage}{0.48\textwidth}
		\centering
		\includegraphics[width=\linewidth]{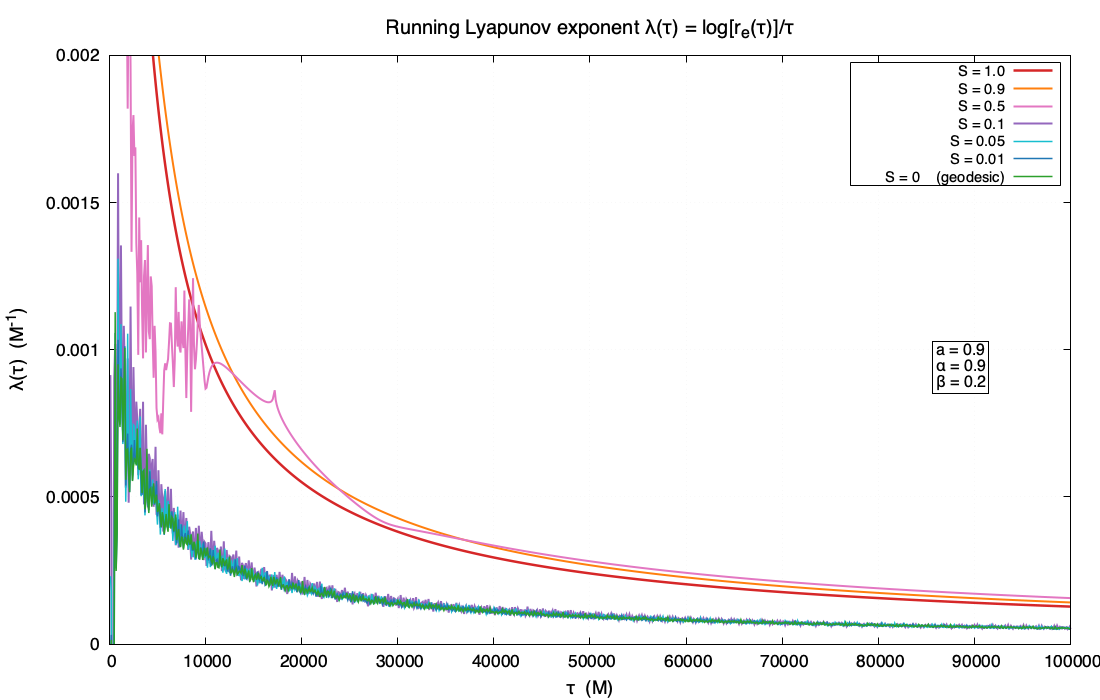}
		\vspace{1mm}
		\centerline{(b) Hairy case: $a=0.9$, $\alpha=0.9$, $\beta=0.2$.}
	\end{minipage}
	\vspace{2mm}
	
	\begin{minipage}{0.48\textwidth}
		\centering
		\includegraphics[width=\linewidth]{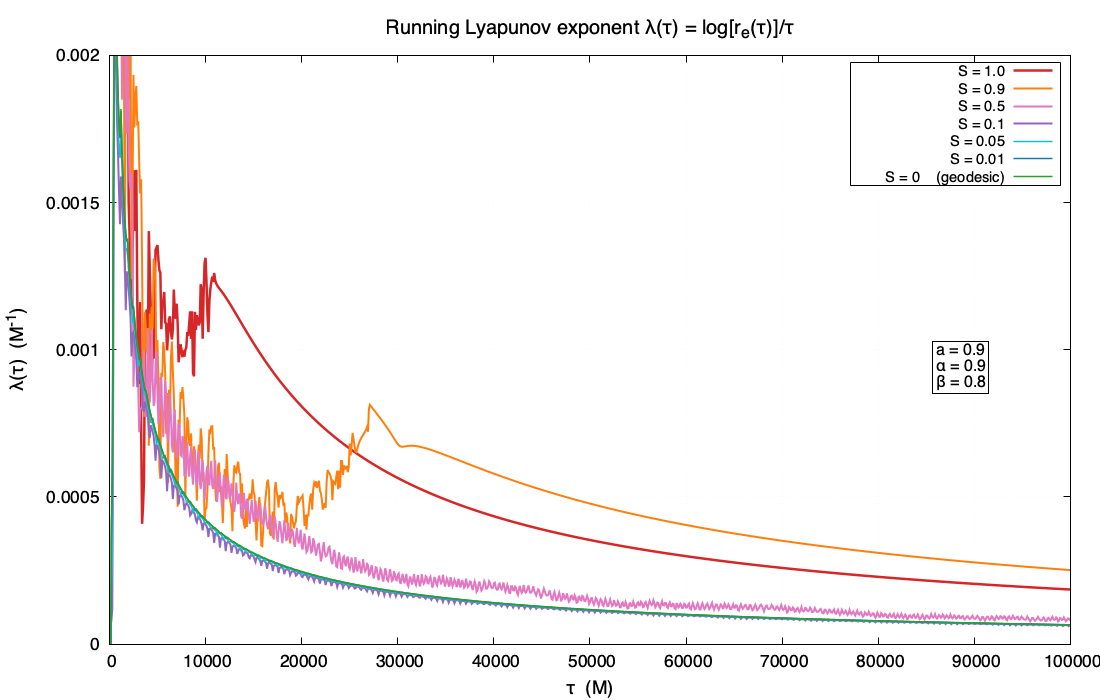}
		\vspace{1mm}
		\centerline{(c) Hairy case: $a=0.9$, $\alpha=0.9$, $\beta=0.8$.}
	\end{minipage}
	\hfill
	\begin{minipage}{0.48\textwidth}
		\centering
		\includegraphics[width=\linewidth]{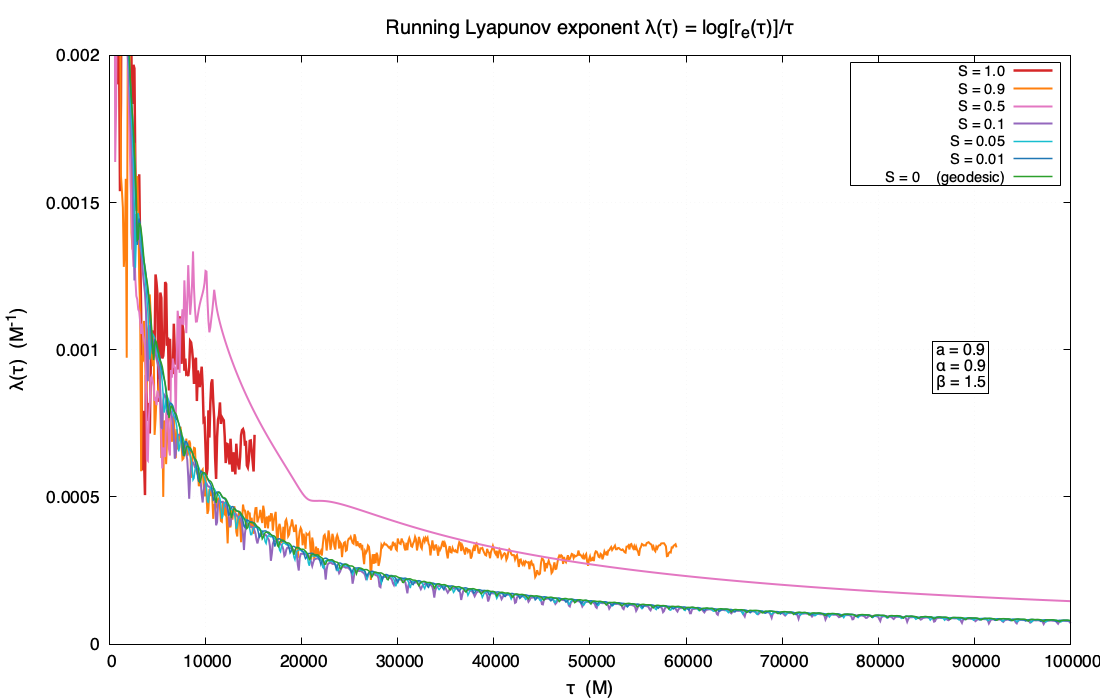}
		\vspace{1mm}
		\centerline{(d) Hairy case: $a=0.9$, $\alpha=0.9$, $\beta=1.5$.}
	\end{minipage}
	
	\caption{Running finite-time Lyapunov exponent
		$\lambda(\tau)=\log[r_e(\tau)]/\tau$ for different values of the
		particle spin $S$.  Panel (a) gives the Kerr limit, since
		$\alpha=0$ switches off the deformation.  Panels (b)--(d) show the
		corresponding rotating hairy black hole cases for different values
		of $\beta$ in the deformed mass function.  The weak-spin
		trajectories, including the geodesic limit $S=0$, remain clustered
		together and decay toward small running exponents.  The high-spin
		trajectories separate from this family and show stronger finite-time
		growth.  The dependence on $\beta$ is non-monotonic, reflecting the
		fact that $\beta$ changes the radial localization of the exponential
		deformation rather than simply increasing the hair strength.  Some
		high-spin curves terminate earlier because the trajectory reaches
		the saturation, plunge, or constraint-violation cutoff before the
		maximum integration time; only the pre-saturation part is used for
		the finite-time Lyapunov estimate.}
	\label{fig:running_lambda_spin_scan}
\end{figure*}

Figure~\ref{fig:running_lambda_spin_scan} summarizes the spin
dependence for the Kerr limit and for three representative hairy
backgrounds.  Panel~\ref{fig:running_lambda_spin_scan}(a) gives the Kerr
limit, obtained by setting $\alpha=0$.  Although the figure label
contains $\beta=0.2$, the deformation term is switched off when
$\alpha=0$, so this panel provides the reference Kerr behavior.  The
geodesic curve $S=0$ and the weak-spin curves $S=0.01$, $0.05$, and
$0.1$ cluster closely together and decay toward small values of
$\lambda(\tau)$.  The high-spin cases are clearly separated from this
regular family.  In particular, $S=1.0$ and $S=0.9$ remain elevated over
a long initial interval before gradually decreasing, while $S=0.5$
shows a stronger transient response than the weak-spin group.  This
confirms that even in the Kerr limit appreciable spin--curvature
coupling is required before a clear finite-time instability appears.

When the hairy deformation is switched on with $\alpha=0.9$
and $\beta=0.2$, shown in panel~\ref{fig:running_lambda_spin_scan}(b),
the same qualitative hierarchy remains, but the separation between the
large-spin and weak-spin families becomes more pronounced.  The curves
with $S=1.0$ and $S=0.9$ dominate the finite-time Lyapunov response,
whereas the small-spin curves remain grouped with the geodesic limit.
The $S=0.5$ trajectory behaves as an intermediate case: it lies above
the weak-spin family over a significant interval, but it does not remain
as strongly separated as the largest-spin trajectories.  This behavior
indicates that the nonzero deformation modifies the curvature region
sampled by the orbit, while the spin magnitude controls how efficiently
the particle responds to that curvature through the MPD force.

Panel~\ref{fig:running_lambda_spin_scan}(c) shows the case
$\alpha=0.9$ and $\beta=0.8$.  Since the deformed mass function
contains the scale $M-\beta/2$ in the exponential factor, changing
$\beta$ changes the radial localization of the hairy deformation rather
than simply increasing its strength.  For this value of $\beta$, the
large-spin trajectories show a more irregular finite-time response.  The
$S=1.0$ curve enters the strong-growth or termination regime early,
while the $S=0.9$ curve exhibits a visible enhancement at intermediate
times before decaying more slowly than the weak-spin family.  The
$S=0.5$ trajectory again occupies a borderline position between the
large-spin sector and the nearly geodesic sector.  The weak-spin and
geodesic curves remain clustered together, indicating that small spin
alone is not sufficient to generate a strong finite-time Lyapunov
signal.

For the more localized deformation with $\alpha=0.9$ and $\beta=1.5$,
shown in panel~\ref{fig:running_lambda_spin_scan}(d), the hierarchy is
again non-monotonic.  The largest-spin trajectory develops a strong
early-time response and does not extend over the full integration time,
because it reaches one of the stopping criteria described in
Sec.~\ref{sec:lyapunov_method}.  The $S=0.9$ curve shows a delayed enhancement
at intermediate times, while $S=0.5$ displays a broad transient before
decaying toward smaller values.  By contrast, the geodesic and
small-spin curves remain close to one another throughout the evolution.
Thus, increasing $\beta$ does not uniformly increase the running
Lyapunov exponent.  Instead, it shifts the radial region in which the
hairy correction is dynamically important, and the finite-time response
depends on whether the evolved MPD trajectory actually samples this
region.

The main physical conclusion from Fig.~\ref{fig:running_lambda_spin_scan}
is therefore twofold.  First, the spin dependence is robust: geodesic
and weak-spin trajectories remain close to a regular family, while
large-spin trajectories display much stronger finite-time sensitivity to
initial conditions.  Second, the effect of the hair is not a simple
monotonic amplification.  The parameter $\beta$ changes the radial
localization of the exponential deformation and thereby reorganizes the
strong-field region sampled by the spinning trajectory.  The largest
running Lyapunov responses occur only when the particle spin is large
enough for spin--curvature coupling to be efficient and when the
empirical orbit probes the part of the geometry where the hairy
deformation is dynamically relevant.

\subsection{Deep integrations and borderline trajectories}
\label{subsec:deep_integrations}

\begin{figure*}[t]
	\centering
	\begin{minipage}{0.48\textwidth}
		\centering
		\includegraphics[width=\linewidth]{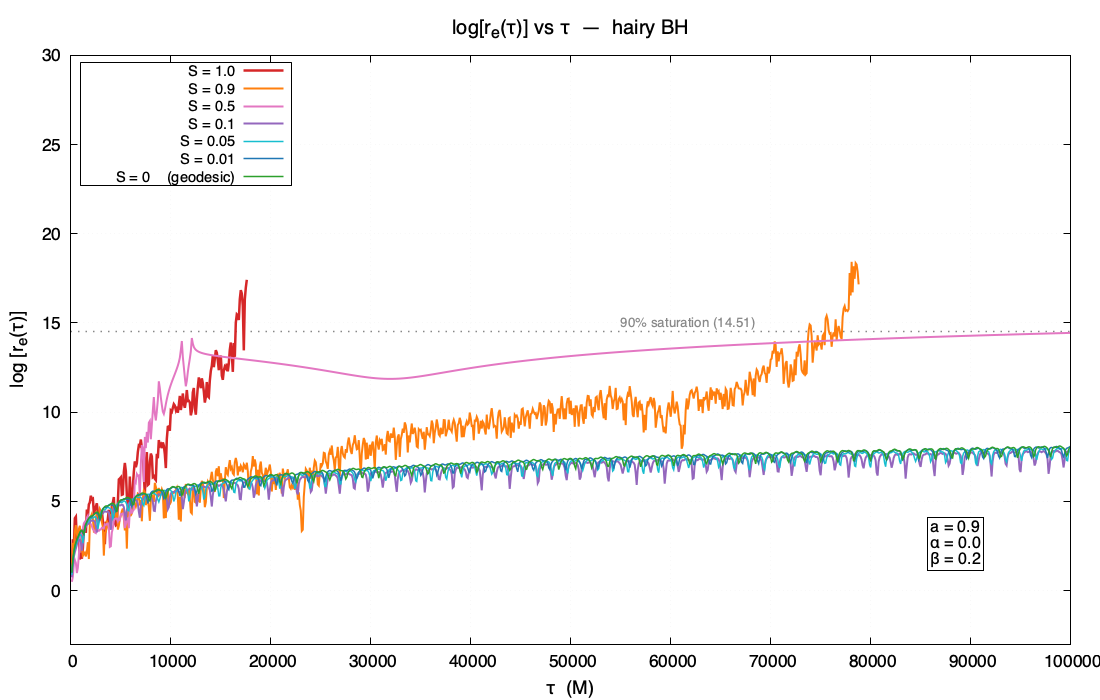}
		\vspace{1mm}
		\centerline{(a) Kerr limit: $a=0.9$, $\alpha=0$, $\beta=0.2$.}
	\end{minipage}
	\hfill
	\begin{minipage}{0.48\textwidth}
		\centering
		\includegraphics[width=\linewidth]{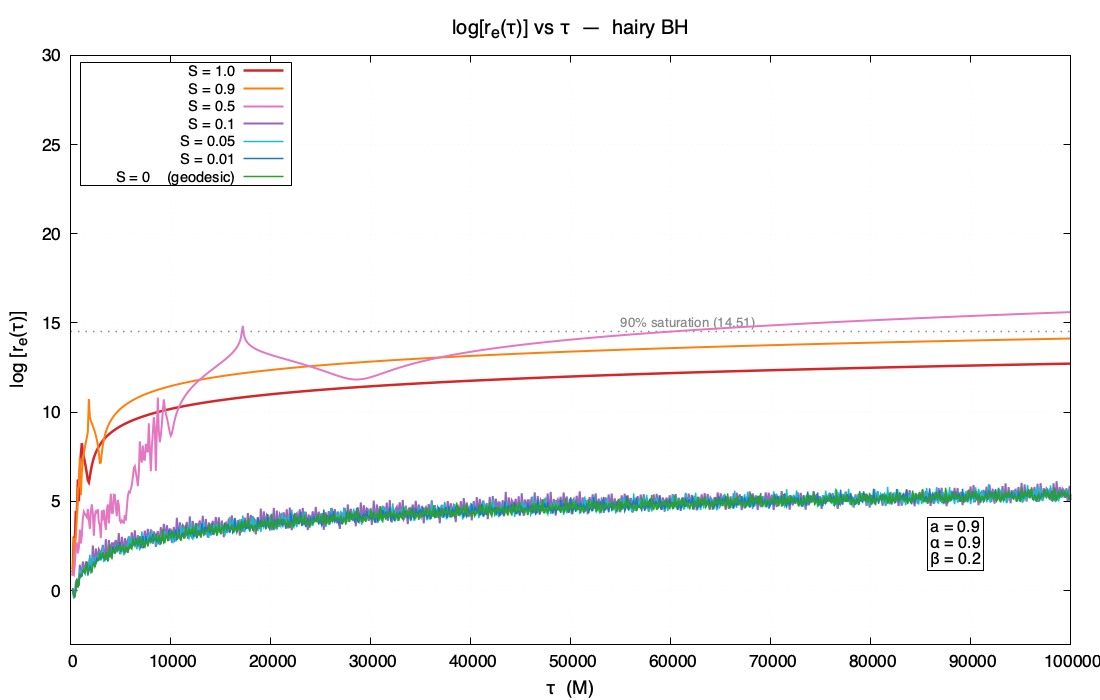}
		\vspace{1mm}
		\centerline{(b) Hairy case: $a=0.9$, $\alpha=0.9$, $\beta=0.2$.}
	\end{minipage}
	\vspace{2mm}
	
	\begin{minipage}{0.48\textwidth}
		\centering
		\includegraphics[width=\linewidth]{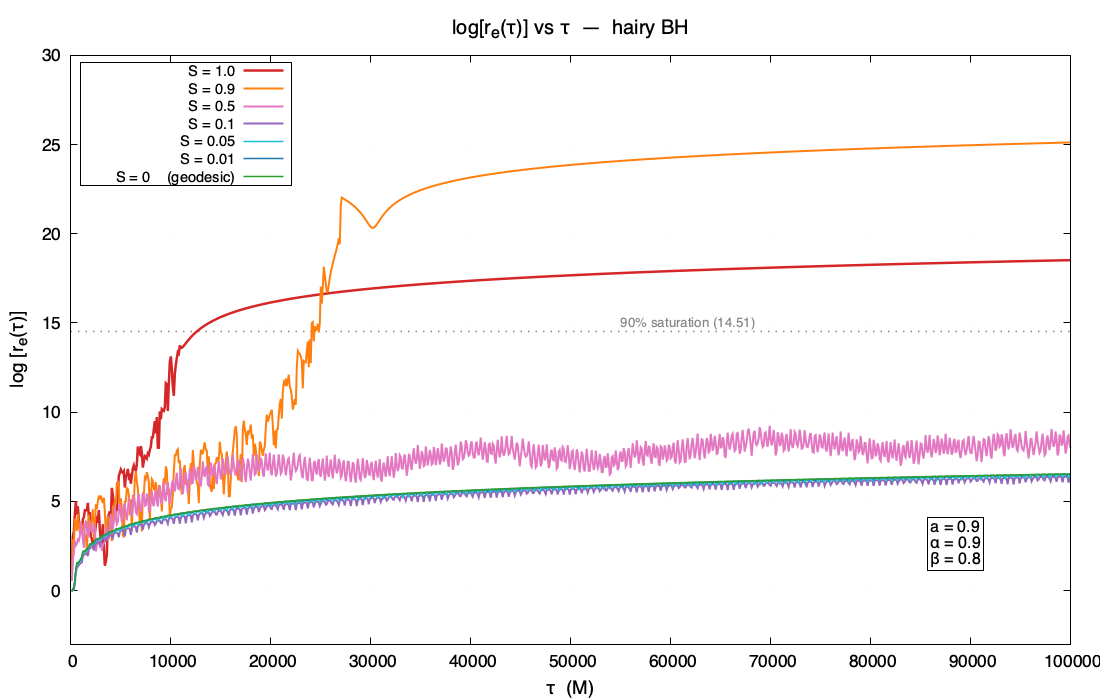}
		\vspace{1mm}
		\centerline{(c) Hairy case: $a=0.9$, $\alpha=0.9$, $\beta=0.8$.}
	\end{minipage}
	\hfill
	\begin{minipage}{0.48\textwidth}
		\centering
		\includegraphics[width=\linewidth]{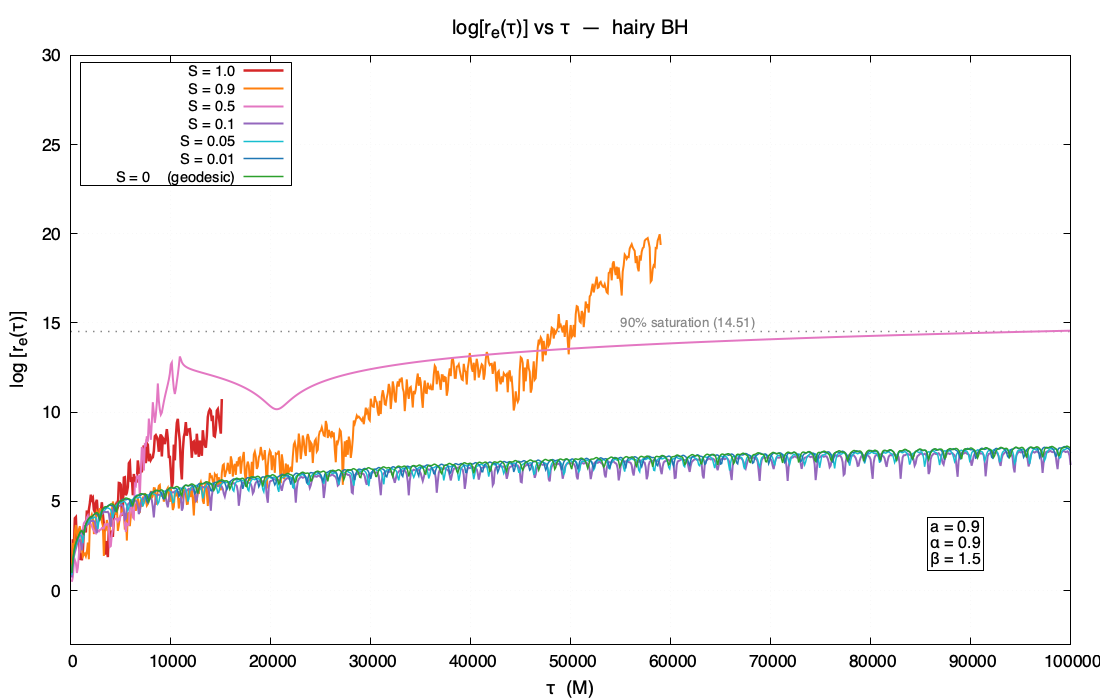}
		\vspace{1mm}
		\centerline{(d) Hairy case: $a=0.9$, $\alpha=0.9$, $\beta=1.5$.}
	\end{minipage}
	
	\caption{Logarithmic growth of the deviation factor
		$\log r_e(\tau)$ for the same spin values and background parameters
		as in Fig.~\ref{fig:running_lambda_spin_scan}.  The horizontal
		dotted line marks the 90\% saturation level of the unrescaled
		deviation-vector method.  Curves that remain well below this line
		correspond to regular or weakly unstable trajectories over the
		integration time, whereas curves that grow steadily toward or beyond
		the line indicate sustained finite-time exponential sensitivity
		before saturation.  The high-spin trajectories generally show the
		strongest growth, while the geodesic and very small spin cases
		remain grouped together.  The relative behavior of the large-spin
		and intermediate-spin curves changes with $\beta$, showing that the
		hairy deformation reorganizes the empirical strong-field
		region sampled by the MPD trajectory rather than simply amplifying
		all Lyapunov indicators uniformly.  Missing late-time portions of
		some high-spin curves indicate that the corresponding trajectory has
		reached the saturation, plunge, or constraint-violation termination
		criterion.}
	\label{fig:log_re_spin_scan}
\end{figure*}

The running exponent $\lambda(\tau)$ is useful for comparing the
relative strength of finite-time growth for different spin values, but
a more direct diagnostic is the logarithmic deviation factor itself.
We therefore examine $\log r_e(\tau)$ for the same representative
backgrounds used in Sec.~\ref{subsec:spin_dependence_running_lambda}.
The horizontal dotted line in Fig.~\ref{fig:log_re_spin_scan} marks the
90\% saturation level of the unrescaled deviation-vector method.  Once a
trajectory approaches or crosses this level, the separation between the
two evolved solutions is no longer infinitesimal, and the later part of
the curve should not be used to extract a local Lyapunov exponent.
Therefore, the physically meaningful Lyapunov estimate is taken only
from the pre-saturation growth window.

Panel~\ref{fig:log_re_spin_scan}(a) shows the Kerr limit,
$\alpha=0$.  The geodesic trajectory and the weak-spin trajectories
remain grouped together and grow only slowly over the full integration
time.  They stay well below the saturation level, indicating that the
corresponding motion is regular or only weakly unstable over the time
interval considered.  The high-spin curves behave very differently.
The $S=1.0$ trajectory grows rapidly and terminates early, indicating
that it reaches the nonlinear saturation or stopping criterion before
the end of the run.  The $S=0.9$ trajectory also grows strongly and
crosses the saturation level at late times.  The intermediate case
$S=0.5$ approaches the saturation line more slowly and should be viewed
as a borderline trajectory.  Thus, even in the Kerr reference case, the
deep integration confirms that strong finite-time separation is tied to
large spin--curvature coupling.

Panel~\ref{fig:log_re_spin_scan}(b) corresponds to the rotating hairy
background with $\alpha=0.9$ and $\beta=0.2$.  In this case the
separation between the weak-spin family and the larger-spin trajectories
is again clear, but the ordering among the large-spin cases is not
identical to the Kerr limit.  The $S=1.0$ and $S=0.9$ curves rise above
the weak-spin/geodesic group and remain well separated from it, while
the $S=0.5$ trajectory shows a particularly strong long-time growth and
eventually crosses the saturation level.  This illustrates an important
point: the spin magnitude alone does not determine the full response.
The hairy deformation changes the empirical strong-field region sampled
by the orbit, and an intermediate spin can therefore become strongly
unstable if the evolved trajectory passes through a favorable curvature
region.

For $\alpha=0.9$ and $\beta=0.8$, shown in
panel~\ref{fig:log_re_spin_scan}(c), the large-spin sector displays the
strongest finite-time growth among the examples shown.  Both $S=1.0$
and $S=0.9$ cross the saturation level and continue into the nonlinear
separation regime.  The $S=0.9$ curve grows especially strongly at
late times, reaching values of $\log r_e(\tau)$ far above the
saturation threshold.  In contrast, the $S=0.5$ trajectory remains
below the saturation level and behaves as an intermediate case, while
the small-spin and geodesic curves remain clustered together.  This
panel gives a clear example of the cooperative spin--hair effect:
large spin provides a strong response to curvature, while the
hairy deformation places the evolved trajectory in a region where that
response is amplified.

Panel~\ref{fig:log_re_spin_scan}(d) shows the more localized
deformation with $\alpha=0.9$ and $\beta=1.5$.  The behavior is again
strongly non-monotonic.  The $S=0.9$ trajectory exhibits a pronounced
late-time growth and crosses the saturation level, whereas the $S=1.0$
trajectory shows strong early growth but terminates earlier.  The
$S=0.5$ case approaches the saturation level only gradually and remains
a borderline trajectory over most of the integration time.  The
geodesic and weak-spin cases remain grouped together and far below the
saturation threshold.  Therefore, increasing $\beta$ does not simply
increase all deviations uniformly.  Since $\beta$ changes the radial
localization scale of the exponential deformation, the resulting
finite-time instability depends on how the empirical trajectory samples
the deformed strong-field region.

These deep integrations also clarify how we distinguish sustained
finite-time instability from transient effects.  Several curves show
bursts, dips, or oscillatory features, especially for intermediate and
large spin.  Such behavior is expected in strongly relativistic
trajectories, where precession, phase mixing, and zoom--whirl-like
episodes can produce temporary enhancements in the deviation factor.
For this reason, we do not classify a trajectory as strongly unstable
solely because of a short burst in $\log r_e(\tau)$.  The relevant
criterion is sustained growth over a sufficiently long pre-saturation
interval, together with the running-exponent behavior discussed in
Sec.~\ref{subsec:spin_dependence_running_lambda}.

The combined message of Figs.~\ref{fig:running_lambda_spin_scan} and
\ref{fig:log_re_spin_scan} is that the geodesic and weak-spin
trajectories form a nearly regular family, whereas large-spin
trajectories can show strong finite-time exponential sensitivity.  The
role of the hairy deformation is not a simple monotonic
amplification of this behavior.  Instead, the hair changes the radial
region over which the curvature is modified and thereby changes which
empirical trajectories encounter the most unstable part of phase space.
This is why the intermediate case $S=0.5$ may behave as a borderline
trajectory in some backgrounds but can become strongly unstable in
others, and why the relative ordering of the $S=1.0$ and $S=0.9$
curves changes as $\beta$ is varied.

\subsection{Combined role of spin and hair}
\label{subsec:combined_spin_hair}

\begin{figure}[t]
	\centering
	\includegraphics[width=1.1\linewidth]{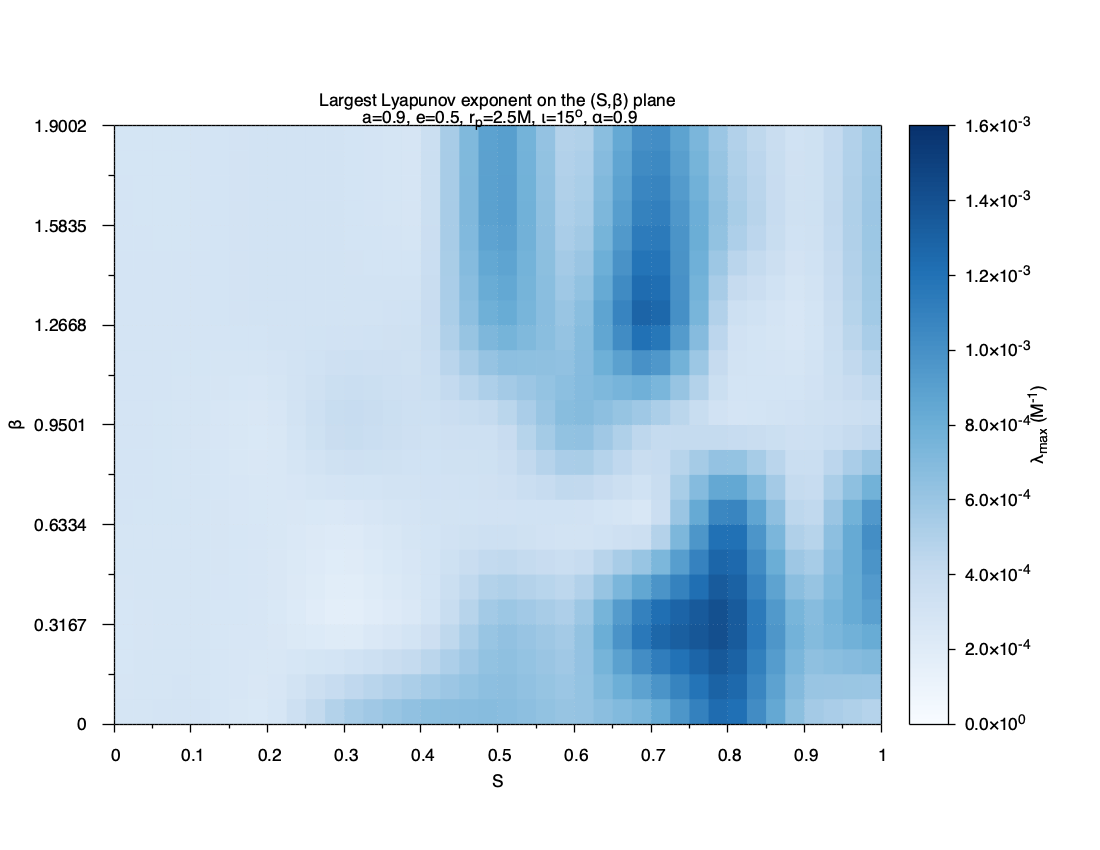}
	\caption{Density plot of the largest finite-time Lyapunov exponent
		$\lambda_{\max}$ in the $(S,\beta)$ plane for fixed
		$a=0.9$, $e=0.5$, $r_p=2.5M$, $\iota=15^\circ$, and
		$\alpha=0.9$.  The color scale gives $\lambda_{\max}$ in units of
		$M^{-1}$.  The instability is not enhanced uniformly throughout
		the parameter space.  Instead, the rotating hairy metric produces
		localized regions of stronger finite-time instability, most
		prominently around $S\simeq 0.75-0.85$, $\beta\simeq 0.2-0.7$, and
		around $S\simeq 0.65-0.75$, $\beta\simeq 1.2-1.5$.  This
		non-monotonic structure shows that the strongest sensitivity to
		initial conditions is governed by the cooperative effect of
		spin--curvature coupling and the radial localization of the hairy
		deformation.}
	\label{fig:S_beta_heatmap_corrected}
\end{figure}

We now examine the simultaneous dependence of the finite-time Lyapunov
indicator on the particle spin and on the hairy deformation.  For this
purpose we scan the $(S,\beta)$ plane while keeping the remaining
orbital and background parameters fixed,
\begin{eqnarray}
	&&a=0.9,\qquad e=0.5,\qquad r_p=2.5M,\nonumber\\
	&&~~~~~~~~~~\iota=15^\circ,\qquad \alpha=0.9 .
\end{eqnarray}
The result is shown in Fig.~\ref{fig:S_beta_heatmap_corrected}.  The
color scale gives the largest finite-time Lyapunov exponent
$\lambda_{\max}$ in units of $M^{-1}$.  This scan is particularly useful
because it separates the role of the particle spin, which controls the
strength of the MPD spin--curvature force, from the role of the hair
parameter, which changes the radial localization of the exponential
deformation in the deformed mass function.

The first important feature of Fig.~\ref{fig:S_beta_heatmap_corrected}
is that the instability is not distributed uniformly over the
parameter space.  For very small spin, the finite-time Lyapunov exponent
remains comparatively weak over most of the $\beta$ range.  This is
consistent with the behavior seen in Figs.~\ref{fig:running_lambda_spin_scan}
and~\ref{fig:log_re_spin_scan}: when $S$ is small, the MPD trajectory
remains close to the geodesic family and the spin--curvature force is
not strong enough to generate large separation of nearby trajectories.
As the spin is increased, $\lambda_{\max}$ grows only in selected
regions of the $(S,\beta)$ plane rather than increasing everywhere.

The rotating hairy metric produces a visibly non-monotonic structure.  The
largest values of $\lambda_{\max}$ appear mainly in two localized
regions.  One enhanced region occurs at intermediate-to-large spin,
approximately
\begin{equation}
	S \simeq 0.75-0.85 ,
\end{equation}
and relatively small-to-moderate values of the hair parameter,
approximately
\begin{equation}
	\beta \simeq 0.2-0.7 .
\end{equation}
A second enhanced region appears around
\begin{equation}
	S \simeq 0.65-0.75 ,
\end{equation}
with
\begin{equation}
	\beta \simeq 1.2-1.5 .
\end{equation}
Between and around these regions the exponent decreases, showing that
the strongest finite-time instability is not controlled by either $S$
or $\beta$ alone.

This structure has a natural interpretation in terms of the MPD force,
\begin{equation}
	F^\mu_{\rm spin}
	=
	-\frac{1}{2}
	R^\mu{}_{\nu\alpha\beta}
	v^\nu S^{\alpha\beta}.
\end{equation}
The spin magnitude determines how efficiently the particle responds to
the curvature, whereas the hairy deformation determines which radial
curvature region is sampled by the empirical trajectory.  In the
deformed mass function, $\beta$ enters through the exponential scale
$M-\beta/2$.  Thus, varying $\beta$ does not simply strengthen the hair
in a monotonic way; it changes where the deformation is radially
localized and therefore changes the strong-field region that the
evolved MPD orbit probes.

For small $S$, the spin--curvature interaction is too weak to produce a
large Lyapunov response even when the background is deformed.  For very
large $S$, the response is strong, but the trajectory can also approach
the saturation, plunge, or constraint-limited regime before a long clean
finite-time growth window is obtained.  Between these limits, the
particle spin and the radial localization of the deformation can become
balanced in such a way that nearby MPD trajectories separate most
efficiently.  This explains why the largest values of
$\lambda_{\max}$ occur in localized bands rather than along the entire
large-spin or large-$\beta$ region.

Figure~\ref{fig:S_beta_heatmap_corrected} also clarifies the role of
the requested-to-empirical map discussed in
Sec.~\ref{subsec:req_emp_maps}.  The point specified by the requested
labels $(r_p,e,\iota)$ does not uniquely determine the physical
strong-field region sampled by the evolved spinning trajectory.  The
hairy deformation can shift the empirical radial and polar behavior,
while the particle spin determines the strength of the response to the
curvature encountered along that empirical orbit.  Therefore, the
enhanced regions in the heatmap should be interpreted as regions where
the empirical trajectory passes through a curvature domain in which the
spin--curvature coupling is particularly efficient.

The main conclusion from the $(S,\beta)$ scan is therefore that the
finite-time instability is a cooperative spin--hair effect.  The
particle spin alone does not determine the strength of the Lyapunov
indicator, and changing the hair parameter alone does not uniformly
increase the instability.  Instead, the rotating hairy geometry
reorganizes the phase-space region sampled by the orbit, and strong
finite-time sensitivity appears only when this reorganization places
the spinning trajectory in a region where the MPD force can efficiently
amplify nearby deviations.

\section{Conclusion}
\label{sec:conclusion}

In this work we have investigated the dynamics of spinning test
particles in a rotating hairy black hole spacetime generated through
gravitational decoupling.  The particle motion was evolved using the
full MPD equations in the spin-one-form
formulation, with the Tulczyjew spin supplementary condition imposed
throughout the numerical evolution.  Instead of reducing the problem to
equatorial circular motion, we followed generic MPD trajectories in the
full phase space and diagnosed their sensitivity to initial conditions
through a ZAMO-projected finite-time Lyapunov analysis.

The first main result is that the distinction between requested and
empirical orbital parameters is essential for interpreting the
dynamics.  The requested labels $(r_p,e,\iota)$ define how the initial
data are generated, but the evolved MPD trajectory need not remain in
the corresponding region of phase space.  Spin--curvature coupling, and
the deformation of the background geometry, shift both the radial
turning points and the polar excursion of the orbit.  As a result, the
same trajectory can appear in rather different locations in the
requested and empirical $(r_p,\iota)$ maps.  The empirical maps are
therefore not merely a diagnostic representation; they identify the
actual strong-field region sampled by the spinning particle.

Compared with the vacuum Kerr MPD problem, the main new effect found
here is the reorganization of the requested-to-empirical phase-space
map by the hairy deformation.  In the rotating hairy metric, the parameter
$\beta$ enters through the exponential radial scale $M-\beta/2$ and
therefore changes the radial localization of the deformation rather
than simply increasing the hair strength monotonically.  This
localization changes the curvature region probed by the empirical
trajectory.  The finite-time Lyapunov response is then determined by
the combination of this empirical orbital reorganization and the
spin--curvature force acting along the trajectory.

The running Lyapunov curves show a robust spin hierarchy.  The geodesic
case and the small-spin trajectories remain close to one another and
show decreasing running exponents over the integration time.  In
contrast, the large-spin trajectories display much stronger finite-time
growth.  This confirms that the enhanced sensitivity is tied to the
spin degree of freedom and to the non-geodesic MPD force.  At the same
time, the comparison between different values of $\beta$ shows that the
background deformation does not uniformly amplify all Lyapunov
indicators.  Instead, changing $\beta$ can enhance, delay, or suppress
the finite-time growth depending on whether the empirical trajectory
passes through the curvature region where the deformation is
dynamically important.

The deep integrations of $\log r_e(\tau)$ support this interpretation.
The weak-spin and geodesic trajectories remain far below the saturation
threshold, whereas the large-spin trajectories can approach or cross
the saturation level after a sustained growth phase.  Some
intermediate-spin trajectories behave as borderline cases, and their
classification depends sensitively on the background parameters.  This
is consistent with the finite-time nature of the Lyapunov analysis:
short bursts in the deviation factor are not by themselves taken as
evidence of chaos.  The relevant signal is sustained pre-saturation
growth, together with the behavior of the running exponent.

The scan in the $(S,\beta)$ plane provides the clearest summary of the
spin--hair interplay.  The largest finite-time Lyapunov exponents are
not distributed uniformly and do not increase monotonically with either
$S$ or $\beta$.  Instead, the rotating hairy metric gives localized regions
of stronger instability, most prominently around
\[
S\simeq 0.75-0.85,\qquad \beta\simeq 0.2-0.7,
\]
and around
\[
S\simeq 0.65-0.75,\qquad \beta\simeq 1.2-1.5 .
\]
These regions indicate that the strongest finite-time instability
appears only when the particle spin and the radial localization of the
hairy deformation are suitably balanced.  The spin controls the
strength of the response to curvature, while the rotating hairy
background controls the curvature region actually sampled by the
evolved MPD trajectory.

We have also emphasized the limitations of the present analysis.  The
large values of the dimensionless spin parameter used in part of the
survey should be interpreted as dynamical probes of the nonlinear MPD
phase space rather than as direct astrophysical EMRI spin values in the
strict pole--dipole regime.  In the large-spin regime, higher multipole
and finite-size effects may become important.  Moreover, the Lyapunov
exponents computed here are finite-time indicators obtained from
pre-saturation growth windows; a complete classification of chaos could
be strengthened further by complementary diagnostics such as Poincare
sections, FLI, SALI, or MEGNO.

Several extensions are natural.  First, a higher-resolution scan of
the $(S,\beta)$ plane would clarify the detailed structure of the
localized instability bands found here.  Second, a comparison with
waveform-level observables would be needed to assess whether the
spin--hair induced changes in the trajectory can produce measurable
dephasing in EMRI-like systems
\cite{Yuan:2026realisticspin,Drasco:2005kz,AmaroSeoane:2017osp,
	Babak:2017tow,Barack:2018yly}.  Third, including higher multipole
moments of the small body would test the robustness of the present
pole--dipole results in the strong spin--curvature regime.  These
directions would help connect the finite-time phase-space signatures
identified here with more directly observable strong-field dynamics.

\begin{acknowledgments}
We would like to thank Rajesh Karmakar for helpful discussions.
This work was partially supported by the National Natural Science
Foundation of China (NSFC) under Grant Nos. 12275166 and 12311540141.
\end{acknowledgments}

\bibliography{Bibliography} 
%%%%%%%%

\end{document}